\documentclass[12pt,a4paper]{article}
\usepackage[a4paper,margin=1in]{geometry}
\usepackage{hyperref}
\usepackage{cite}
\usepackage{units}
\usepackage{epsfig}
\usepackage{pifont}
\usepackage{amsmath,amsfonts,amssymb,amsthm,mathrsfs}
\usepackage{graphicx,chngcntr,slashed}
\usepackage{cancel}
\usepackage{afterpage}
\usepackage{braket}
\usepackage{caption}
\usepackage{subcaption}
\usepackage{xcolor}
\usepackage{makecell}
\usepackage{physics}
\usepackage{multirow}
\usepackage{float}
\usepackage[normalem]{ulem}

\newcommand{\be}{\begin{equation}}
\newcommand{\ee}{\end{equation}}
\newcommand{\bea}{\begin{eqnarray}}
\newcommand{\eea}{\end{eqnarray}}
\def\ww#1{\langle #1 \rangle}

\newcommand{\ct}{{\rm cos}(\theta)}
\newcommand{\st}{{\rm sin}(\theta)}

\makeatletter
\def\section{\@startsection {section}{1}{\z@}{+3.0ex plus +1ex minus
  +.2ex}{2.3ex plus .2ex}{\large\bf\boldmath}}
\def\subsection{\@startsection{subsection}{2}{\z@}{+2.5ex plus +1ex
minus +.2ex}{1.5ex plus .2ex}{\normalsize\bf\boldmath}}
\def\subsubsection{\@startsection{subsubsection}{3}{\z@}{+3.25ex plus
 +1ex minus +.2ex}{1.5ex plus .2ex}{\normalsize\it}}

\graphicspath{{.}{plots/}}

\begin{document}
\thispagestyle{empty}

\def\thefootnote{\fnsymbol{footnote}}

\begin{flushright}
\end{flushright}

\vspace{1cm}

\begin{center}

{\Large {\bf Single-spin measurements and heavy new physics in the $e^+e^- \to t\bar{t}$ process at an FCC-ee}}
\\[3.5em]

{\large
Haotian Cao$^{1,2}$, Frank~Petriello$^1$
}

\vspace*{1cm}

{\sl
$^1$ Department of Physics \& Astronomy, Northwestern University,\\ Evanston, Illinois 60208, USA\\
$^2$ Joint BNL-SBU Center for Frontiers in Nuclear Science (CFNS),\\
Stony Brook University, Stony Brook, New York 11794, USA\\
Email: \href{mailto:haotiao.cao@northwestern.edu}{\texttt{haotiao.cao@northwestern.edu}},\;
\href{mailto:f-petriello@northwestern.edu}{\texttt{f-petriello@northwestern.edu}}
}

\end{center}


\begin{abstract}
    
We investigate the potential of single-spin components of the
spin-density matrix in the $e^+
e^- \to t\bar{t}$ process at a future FCC-ee for probing heavy new
physics parametrized using the SMEFT framework. We consider the full
spectrum of spin observables and the complete angular decomposition of the $t\bar{t}$
production process in our study. We find that single-spin measurements 
generically provide stronger probes of SMEFT Wilson coefficients than measurements where the
$t\bar{t}$ spins are correlated, and that single-spin observables are
important for resolving flat directions that can appear in the Wilson-coefficient parameter space.
    
\end{abstract}

\setcounter{page}{0}
\setcounter{footnote}{0}


\section{Introduction}

The top quark is the heaviest known elementary fermion, with a Yukawa coupling to the Higgs boson of approximately unity. This unique property has long led physicists to believe that it may play a special role in electroweak symmetry breaking, and that measurements of its properties may help us better understand the Higgs sector of the Standard Model(SM) and its possible extensions. Measurements of top-quark properties have been actively pursued since its discovery at the Tevatron experiments decades ago~\cite{CDF:2010yag,D0:2011kcb,D0:2011rkb,D0:2012wfp}. The top quark mass and many of its gauge couplings have been determined to high accuracy through measurements at the LHC~\cite{Miralles:2021dyw}.

A special feature of the top quark that enables a detailed study of its connection to potential new physics is its short lifetime. It is the only known quark that decays prior to hadronization. The properties of the top quark, such as its spin, are imprinted on the bottom quark and the $W$ boson produced in its decay. These properties can be accessed through measurements of the decay-product angular distributions. There have been detailed experimental studies of how top-quark spin properties manifest themselves in the lepton angular distributions arising from $W$-boson decay~\cite{ATLAS:2023fsd,CMS:2024pts,ATLAS:2014aus,CMS:2015cal,CMS:2016piu,ATLAS:2016bac,ATLAS:2019hau,ATLAS:2019zrq,CMS:2019nrx,CMS:2024zkc}, and how new physics may affect these properties. These advances have opened new possibilities for exploring quantum observables at the TeV scale and have motivated further phenomenological investigations into the quantum aspects of top-quark pair production~\cite{Maltoni:2024tul,Barr:2024djo,Aoude:2022imd,Severi:2022qjy,Fabbrichesi:2024xtq,Fabbrichesi:2022ovb,Altakach:2022ywa,Aoude:2023hxv,Bernal:2023ruk,Fabbrichesi:2023jep}.

The standard framework used to investigate top-quark spin properties is the spin-density matrix formalism in $t\bar{t}$ production. The formalism for describing the spin structure of the $t\bar{t}$ cross section has been extensively developed for hadronic collisions such as at the LHC~\cite{Aoude:2022imd,Dong:2023xiw,Fabbrichesi:2021npl,Severi:2021cnj,Aguilar-Saavedra:2022uye,Fabbrichesi:2022ovb,Afik:2022dgh,Severi:2022qjy,Aguilar-Saavedra:2023hss,Han:2023fci,Aguilar-Saavedra:2024hwd,Maltoni:2024csn,Brandenburg:2002xr,Bernreuther:2004jv,Bernreuther:2015yna, Bernreuther:2024ltu}. A recent pedagogical overview is given in~\cite{Barr:2024djo}. Significant work has also been done to understand the spin-density matrix in $e^+e^-$ collisions~\cite{Khiem:2015ofa}. This case is particularly interesting as the production of top-quark pairs is driven by its electroweak couplings in $e^+e^-$ collisions rather than its strong coupling as at the LHC. QCD corrections to spin observables have been shown to be small, allowing a simple theoretical interpretation of the measurements~\cite{Brandenburg:1998xw}. Previous work has established that the Future Circular Collider with $e^+e^-$ collisions (FCC-ee) can serve as a powerful probe of the electroweak top-quark vertex~\cite{Janot:2015yza} and possible anomalous electron-top contact interactions~\cite{Bellafronte:2025ubi}. These considerations have motivated further exploration of top-quark properties and couplings at future lepton colliders beyond the HL-LHC era~\cite{Vryonidou:2018eyv,Durieux:2018ekg,Durieux:2018ggn,Durieux:2018tev,Durieux:2019rbz,Jung:2020uzh,Bernardi:2022hny,deBlas:2022ofj,Durieux:2022cvf,Banelli:2020iau}. Spin measurements have been revisited recently in the context of quantum entanglement between the top and anti-top spin states, both at the LHC~\cite{Afik:2020onf,Maltoni:2024tul} and at a future FCC-ee~\cite{Maltoni:2024csn}. In~\cite{Maltoni:2024csn}, a detailed study of how SMEFT operators are constrained by various quantum entanglement measures at an FCC-ee was performed, with some discussion of how these measurements can help remove flat directions in the Wilson coefficient parameter space present in inclusive cross section measurements.
 
In this manuscript, we discuss the sensitivity of FCC-ee to new physics in the top-quark sector. We consider the full spectrum of spin measurements, inclusive cross sections, forward-backward asymmetries, and the full angular decomposition of the $t\bar{t}$ production process at a future FCC-ee collider. We model possible new physics using the Standard Model Effective Field Theory (SMEFT) and include all operators that affect the production of a $t\bar{t}$ pair at tree-level. Top physics at a future FCC-ee will feature a scan of the $t\bar{t}$ production threshold with center-of-mass energies ranging from 345 to 365 GeV, and we study what probes are possible at different collider energies. We analytically study the matrix elements in the threshold limit and highlight the information that can be extracted from measurements of each coefficient in the spin density matrix. A focus of our study is what combinations of measurements are needed to remove flat directions in the Wilson coefficient space. Building upon previous analyses of spin-correlated observables~\cite{Maltoni:2024csn}, we emphasize the importance of single-spin components of the density matrix and forward-backward asymmetries for this purpose. We additionally show that single-spin measurements can provide the strongest single-parameter bounds among the observables we study for several Wilson coefficients.

Our paper is organized as follows. We review the spin-density matrix formalism and present both the SM and SMEFT calculations of its components in Section~\ref{sec:formalism}. In Section~\ref{sec:setup} we discuss the parameters for our scan of our FCC-ee $t\bar{t}$ threshold study and define the observables we consider. We present our numerical results for the expected probes of the SMEFT parameter space in Section~\ref{sec:numerics}. We conclude in Section~\ref{sec:conc}. In an Appendix we list several formulae and numerical results that are not needed for the main discussion in the text.

\section{Formalism}
\label{sec:formalism}

We begin by discussing the spin-density matrix for $t\bar{t}$ production in $e^+ e^-$ collisions. At leading order in perturbation theory, the process $e^+e^- \to t\bar{t}$ proceeds through $s$-channel exchange of a photon or $Z$-boson. We will later consider top-quark decays leading to either a di-lepton or lepton+jet final state. In the narrow-width limit for the top quarks, we can express the full cross section for the di-lepton and lepton+jet final states as the production of a $t\bar{t}$ pair with each top quark having a definite spin, times the differential decay widths for the polarized top quarks. Detailed analytic expressions for the top-quark differential decay widths have been presented in the literature~\cite{Bernreuther:2015yna}. QCD corrections to the differential decay widths are small~\cite{Brandenburg:1998xw}, and are neglected in our study. To obtain the correct inclusive cross section for the $e^+e^- \to t\bar{t}$ process in our numerical results, we normalize our integrated leading-order cross section to the result obtained using {\tt QQbar\_threshold}~\cite{Beneke:2016kkb}, which includes the known higher-order corrections. 

The matrix-element squared for $e^+e^- \to t\bar{t}$ can be written as follows, assuming that the top and anti-top have spin directions $\hat{s}_t$  and $\hat{s}_{\bar{t}}$ respectively:
\be
|{\cal M}(\hat{s}_t ,\hat{s}_{\bar{t}} )|^2 = A + \hat{s}_t B^{(t)} + \hat{s}_{\bar{t}} B^{(\bar{t})} +\hat{s}_t  \hat{s}_{\bar{t}}  C^{(t\bar{t})} .
\label{eq:MEdef}
\ee
Here $A$ represents the unpolarized result, while $C^{(t\bar{t})}$ determines the correlation between the top and anti-top spins. The single-spin coefficients $B^{(t)} $ and $B^{(\bar{t})}$ indicate the terms in which the final state top or antitop is polarized, while the other fermion spin is summed over.

It is convenient to choose a basis for the top quark spins. We denote the top-quark direction of motion as $\hat{k}$, the positron direction of motion as $\hat{p}$, and the angle between these two directions as $\theta$: $\hat{k} \cdot \hat{p} = \ct$ (we abbreviate $\ct = c_t$ and $\st=s_t$ henceforth). From these, we construct the following two vectors:
\be
\hat{r} = \frac{\hat{p} - \hat{k}\, c_t}{s_t}, \;\;\; \hat{n} = \frac{\hat{p} \times \hat{k}}{s_t}.
\ee
We choose $\hat{k}$, $\hat{r}$, and $\hat{n}$ as our basis for the top and anti-top spins. It is straightforward to derive the leading-order expressions in the SM for each of the terms appearing in the matrix elements. In addition to the angle $\theta$, the matrix elements depend upon the top-quark velocity, which we write as 
\be
\beta = \sqrt{1-\frac{4m_t^2}{s}}.
\ee
Following standard conventions in the literature, we split $A$, $B^{(t)}$, $B^{(\bar{t})}$ and $C^{t\bar{t}}$ into coefficient functions and spin-dependent terms.

\subsection{Standard Model results}
\label{sec:SM}

We begin by presenting the SM results for the various components of the matrix elements, in order to establish our conventions. These results have been presented previously in~\cite{Maltoni:2024csn}.

\begin{itemize}

\item {\bf Unpolarized terms}: we can express the SM in the form
\be
A_{[SM]} = F^0_{[SM]} (2-\beta^2+\beta^2 c_t^2 ) +F^1_{[SM]} c_t +F^{At}_{[SM]} (1+c_t^2).
\label{eq:incFdep}
\ee
The coefficient functions $F^i$ have simple expressions, which are collected in the Appendix. The $F^{At}$ terms contains terms with two powers of the top-quark axial coupling $g_{at}$. As we will be interested in the threshold limit since most of the FCC-ee data is expected to be taken near its vicinity, we note the leading scaling of these terms with the top velocity $\beta$:
\be
F^0 \sim 1,\;\;\; F^1 \sim \beta, \;\;\; F^{At} \sim \beta^2.
\ee

\item {\bf Single-spin terms}: there are three single-spin coefficients depending on whether the top quark spin is along the $\hat{k}$, $\hat{r}$, or $\hat{n}$ direction. We denote these as $B^{(t)}_i$ for the index $i$ being one of these three choices. We note that the top and anti-top single-spin coefficients are identical for all choices of spin: $B^{(t)}_i = B^{(\bar{t})}_i$. The coefficient $B^{(t)}_n=0$ at leading order, due to the lack of absorptive amplitude contributions at this order~\cite{Bernreuther:2015yna}. This leaves us with two single-spin coefficients to compute, which we write as
\bea
B^{(t)}_{k,[SM]} &=& (1+c_t^2) G^0_{[SM]}  + c_t \left[ G^1_{[SM]}  + G^{At}_{[SM]}  \right]  \, ,\nonumber \\
B^{(t)}_{r,[SM]} &=& \sqrt{1-\beta^2} c_t s_t G^0_{[SM]}  + \sqrt{1-\beta^2} s_t G^1_{[SM]}  \, .
\label{eq:Bangles}
\eea
The $G^i_{[SM]} $ coefficients are given in the Appendix. The functional dependence of these coefficients on the top-quark velocity is:
\be
G^0 \sim \beta,\;\;\; G^1 \sim 1, \;\;\; G^{At} \sim \beta^2.
\ee
We note that the $G_i$ depend on different combinations of SM couplings than their $F^i$ counterparts, as can be seen from the expressions in the Appendix. This will be important when we move to new physics contributions in the SMEFT.

\item {\bf Spin-correlated terms}: these form a 3-by-3 symmetric matrix written as $C^{(t\bar{t})}_{ij}$. The $C_{rn}$, $C_{kn}$ terms vanish at leading order, again due to the lack of absorptive amplitude contributions~\cite{Bernreuther:2015yna}. This leaves us with four independent quantities that we write as 
\bea
C^{(t\bar{t})}_{kk,[SM]} &=& F^0_{[SM]} (\beta^2+(2-\beta^2) c_t^2 ) +F^1_{[SM]} c_t +F^{At}_{[SM]} (1+c_t^2)\, ,\nonumber \\
C^{(t\bar{t})}_{rr,[SM]} &=& F^0_{[SM]} (2-\beta^2) s_t^2  -F^{At}_{[SM]} s_t^2\, ,\nonumber \\
C^{(t\bar{t})}_{nn,[SM]} &=& -F^0_{[SM]} \beta^2 s_t^2 +F^{At}_{[SM]} s_t^2\, ,\nonumber \\
C^{(t\bar{t})}_{kr,[SM]} &=& 2F^0_{[SM]} \sqrt{1-\beta^2} c_t s_t + F^1_{[SM]} \frac{\sqrt{1-\beta^2}}{2} s_t \, .
\eea
We note that the spin-correlated terms depend on the same coupling-constant combinations $F^i$ as the unpolarized result, unlike the single-spin coefficients. The discussion of these terms is often framed in terms of quantum entanglement~\cite{Afik:2020onf,Maltoni:2024csn}, although whether such measurements are actually a probe of quantum entanglement has been the subject of recent debate~\cite{Abel:2025skj}.

\end{itemize}

\subsection{SMEFT results}

The primary purpose of this paper is to see how well the FCC-ee can probe heavy new physics, which we parameterize using the SMEFT. The SMEFT is an effective field theory extension of the SM that includes higher-dimensional operators suppressed by an energy scale $\Lambda$. Beyond this scale the ultraviolet completion of the EFT 
becomes important, and new particles beyond the SM appear. The SMEFT maintains all gauge symmetries of the SM. In our study, we keep terms through dimension-6 in the $1/\Lambda$ expansion, and ignore operators of odd-dimension which violate lepton number. Our Lagrangian becomes
\begin{equation}
{\cal L} = {\cal L}_{SM}+ \frac{1}{\Lambda^2} \sum_i C^{(6)}_{i} {\cal O}^{(6)}_{i} + \ldots \, ,
\end{equation}
where the ellipsis denotes operators of higher dimensions. The Wilson coefficients $C^{(6)}_{i}$ are dimensionless. Cross sections computed through linear order in the Wilson coefficients will have interferences between dimension-6 operators and the SM. Since our study focuses on the threshold region at a future FCC-ee we will mostly focus on the interference terms that scale as $1 / \Lambda^2$, as the $1 / \Lambda^4$ terms are numerically small for most of our results. We check this assumption explicitly for several cases when we present our numerical results. We use the Warsaw basis~\cite{Grzadkowski:2010es} to parametrize the dimension-6 effects considered in our study.

We will focus here on operators that affect top-quark production. Possible new physics corrections to the top-quark decay vertex $tbW$ have been constrained by LHC measurements~\cite{Buckley:2015lku,Boughezal:2019xpp}. There are three categories of operators needed in our study: four-fermion contact interactions between the electron and top quark; corrections to the top quark vertex interaction with the $Z$-boson, and electroweak dipole operators. We summarize these operators below.
\begin{itemize}

\item {\bf Four-fermion interactions}:
\bea
{\cal O}_{Ql}^{(1)} &=& (\bar{Q} \gamma^{\mu} Q )(\bar{l} \gamma_{\mu} l )\, ,\nonumber \\
{\cal O}_{Ql}^{(3)} &=& (\bar{Q} \gamma^{\mu} \tau^I Q )(\bar{l} \gamma_{\mu}  \tau^I l )\, ,\nonumber \\
{\cal O}_{Qe} &=&  (\bar{Q} \gamma^{\mu} Q )(\bar{e} \gamma_{\mu} e )\, ,\nonumber \\
{\cal O}_{tl} &=&  (\bar{t} \gamma^{\mu} t )(\bar{l} \gamma_{\mu} l )\, ,\nonumber \\
{\cal O}_{te} &=& (\bar{t} \gamma^{\mu} t )(\bar{e} \gamma_{\mu} e ).
\eea
Here, $Q, l$ denote the left-handed SU(2) doublets while $t,e$ denote right-handed SU(2) singlets. The $\tau^I$ are the Pauli sigma matrices that act on the SU(2) indices. It is more convenient to work with the vector and axial combinations of operators rather than the left and right-handed ones used in the Warsaw basis, so we define the following four linear combinations of Wilson coefficients that appear in our expressions:
\bea
C_{VV} &=& \frac{1}{4} \left( C_{Ql}^{(1)} - C_{Ql}^{(3)}  + C_{te}+C_{tl}+C_{Qe} \right) \, ,\nonumber \\
C_{AV} &=& \frac{1}{4} \left( -(C_{Ql}^{(1)} - C_{Ql}^{(3)})  + C_{te}+C_{tl}-C_{Qe} \right) \, ,\nonumber \\
C_{VA} &=& \frac{1}{4} \left( -(C_{Ql}^{(1)} - C_{Ql}^{(3)}) + C_{te}-C_{tl}+C_{Qe} \right) \, ,\nonumber \\
C_{AA} &=& \frac{1}{4} \left( C_{Ql}^{(1)} - C_{Ql}^{(3)}  + C_{te}-C_{tl}-C_{Qe} \right) \, .\nonumber \\
\eea

\item {\bf Top-$Z$ vertex corrections}: These operators affect the $t\bar{t}Z$ electroweak vertex. We note that ${\cal O}_{HQ}^{(3)}$ also affects the $tbW$ vertex that mediates top-quark decay in the SM. While we list these operators here for completeness and include them in our analytic results, we do not study them numerically.
\bea
{\cal O}_{HQ}^{(1)} &=& i (H^{\dagger} \overset{\leftrightarrow} {D_{\mu}} H) (\bar{Q} \gamma^{\mu} Q) \, ,\nonumber \\
{\cal O}_{HQ}^{(3)} &=&  i (H^{\dagger} \overset{\leftrightarrow} {D_{\mu I}} H) (\bar{Q} \gamma^{\mu} \tau^I Q) \, ,\nonumber \\
{\cal O}_{Ht}&=&i (H^{\dagger} \overset{\leftrightarrow} {D_{\mu}} H) (\bar{t} \gamma^{\mu} t) \, .
\eea
Similar to the four-fermion sector, we introduce the following combinations of Wilson coefficients to make the chiral structure explicit:
\begin{align}
    &C_{H Q}^{(3)} + C_{H Q}^{(1)}, \\
    &C_{HV} = \frac{1}{2} \big( C_{H t} + C_{HQ}^{(1)} - C_{HQ}^{(3)}\big), \\
    &C_{HA} = \frac{1}{2} \big( C_{H t} - C_{HQ}^{(1)} + C_{HQ}^{(3)}\big).
\end{align}
We note that $C_{H Q}^{(3)} + C_{H Q}^{(1)}$ does not contribute to $t \bar t$ production at leading order. 
\item {\bf Electroweak dipole operators}: We also include electroweak dipole operators that can modify top-quark production at an FCC-ee.
\bea
{\cal O}_{tW} &=& (\bar{Q} \sigma^{\mu\nu} \tau^I t) \tilde{H} W^{I}_{\mu\nu}\, , \nonumber \\ 
{\cal O}_{tB} &=& (\bar{Q} \sigma^{\mu\nu} t) \tilde{H} B_{\mu\nu} \, .
\eea
The operators $\mathcal O_{tW}$ and $\mathcal O_{tB}$ modify electroweak top-pair production, while $\mathcal O_{tW}$ also affects the $tW^+b$ vertex. Since the spin information of the top quarks is preserved in their decays even in the presence of electroweak dipoles~\cite{Severi:2022qjy}, we include dipole insertions only in the $t\bar t$ production vertex. After electroweak symmetry breaking, it is convenient to define the corresponding Wilson coefficients as follows:
\bea
    C_{tZ} &= c_w C_{tW} - s_w C_{tB}, \\
    C_{t\gamma} &= s_w C_{tW} + c_w C_{tB}.
\eea
The modified $t\bar t Z$ vertex depends only on $C_{tZ}$, while the modified $t\bar t\gamma$ vertex depends only on $C_{t\gamma}$. 

\end{itemize}

For both the four-fermion interactions and vertex corrections no new spin structures are generated beyond those already present in the SM. The dipole operators do generate new spin structures. For the four-fermion operators and vertex corrections, the effect is to extend the coefficients defined in Section~\ref{sec:SM}:
\bea
F^i &=& F^i_{[SM]} +\frac{1}{\Lambda^2}F^i_{[d6]} +\frac{1}{\Lambda^4}F^i_{[d8]} \, ,\nonumber \\
G^i &=& G^i_{[SM]} +\frac{1}{\Lambda^2}G^i_{[d6]} +\frac{1}{\Lambda^4}G^i_{[d8]}\, .
\eea
The $d6$ and $d8$ subscripts denote the $1 / \Lambda^2$ and  $1 / \Lambda^4$ contributions, respectively. We give the $d6$ coefficients below in order to discuss aspects of them, while we relegate the $d8$ ones to the Appendix.
\bea
F^0_{[d6]} &=& 8 N_C e^2 Q_e Q_t C_{VV}\frac{m_t^2}{1-\beta^2} + 
	 32 N_C  \frac{m_t^4}{(1-\beta^2) D_Z} g_{vt} \left\{ g_{ve}C_{VV} +g_{ae} C_{VA} \right\}\nonumber \\
     &-&\frac{4N_C e^3 Q_tQ_e C_{H V} g_{ve} v^2 m_t^2}{c_w s_w D_Z}-\frac{16 N_C e C_{H V} g_{vt} v^2  m_t^4 \left(g_{ae}{}^2+g_{ve}{}^2\right)}{c_w s_w D_Z{}^2} \, , \nonumber \\	
F^{1}_{[d6]} &=& 16 N_C e^2 Q_e Q_t C_{AA} \beta \frac{m_t^2}{1-\beta^2} + 64 N_C \beta \frac{m_t^4}{(1-\beta^2) D_Z} g_{vt} \left\{ g_{ve} C_{AA}+g_{ae} C_{AV}\right\} \nonumber \\
	&+& 64 N_C  \beta \frac{m_t^4}{(1-\beta^2) D_Z} g_{at} \left\{ g_{ve} C_{VA}+g_{ae} C_{VV}\right\} -\frac{8N_C Q_tQ_e\beta  e^3 C_{HA} g_{ae} v^2 m_t^2}{c_w s_w D_Z}
    \nonumber \\
    &-&\frac{64 N_C\beta  e^2 g_{ae} g_{ve} v^2 m_t^4 (C_{H A} g_{vt}+C_{H V} g_{at})}{c_ws_w D_Z^2  }\, ,\nonumber \\
F^{At}_{[d6]} &=& 32 N_C  \frac{\beta^2 m_t^4}{(1-\beta^2) D_Z} g_{at} \left\{ g_{ve} C_{AV}+g_{ae} C_{AA}\right\}-\frac{16N_C \beta ^2 e C_{HA} g_{at} v^2 m_t^4 \left(g_{ae}{}^2+g_{ve}{}^2\right)}{c_w s_w D_Z{}^2 }\, . \nonumber \\
\label{eq:Fid6}
\eea
We have written the $Z^{\mu}\bar{f}f$ vertices appearing in the SM in the form $i \gamma^{\mu}(g_{vf} +g_{af} \gamma^5)$, and have introduced the abbreviation $D_Z = 4 m_t^2 -M_Z^2(1-\beta^2)$ for the $Z$-boson propagator contribution. Note that the 
$F^{At}_{[d6]} $ term contains two axial couplings, one from the SM and one from the new physics interactions. The single-spin coefficients can be expressed as:
\bea
G^0_{[d6]} &=& 8 N_C e^2 Q_e Q_t \beta C_{AV}\frac{m_t^2}{1-\beta^2} + 
	 32 N_C \beta \frac{m_t^4}{(1-\beta^2) D_Z} g_{vt} \left\{ g_{ve} C_{AV}+g_{ae} C_{AA}\right\}  \nonumber \\	
	 &+& 32 N_C\beta \frac{m_t^4}{(1-\beta^2) D_Z} g_{at} \left\{ g_{ve} C_{VV}+g_{ae} C_{VA}\right\}-\frac{4N_C Q_t Q_e\beta  e^3 C_{HA} g_{ve}v^2 m_t^2}{c_w s_w D_Z} \nonumber \\
     &-&\frac{16 N_C \beta  ev^2 m_t^4 \left(g_{ae}{}^2+g_{ve}{}^2\right) (C_{HA} g_{vt}+C_{HV} g_{at})}{c_w s_w D_Z^2} \, ,\nonumber \\
G^{1}_{[d6]} &=& 16 N_C e^2 Q_e Q_t C_{VA} \frac{m_t^2}{1-\beta^2} + 64 N_C \frac{m_t^4}{(1-\beta^2) D_Z} g_{vt} \left\{ g_{ve} C_{VA}+g_{ae} C_{VV}\right\} \nonumber \\
&-&\frac{64 N_C e g_{ae} g_{ve}v^2 m_t^4 C_{HV} g_{vt}}{c_w s_wD_Z{}^2}-\frac{8N_C Q_tQ_ee^3 C_{HV} g_{ae}v^2 m_t^2}{c_w s_w D_Z}\, ,\nonumber \\
G^{At}_{[d6]} &=& 64 N_C \frac{\beta^2 m_t^4}{(1-\beta^2) D_Z} g_{at} \left\{ g_{ve} C_{AA}+g_{ae} C_{AV}\right\}-\frac{64 N_C e g_{ae} g_{ve}v^2 m_t^4 \beta ^2 C_{HA} g_{at}}{c_w s_wD_Z{}^2}\, .
\eea

Before performing numerical studies, we first gain analytical insight into the information accessible from measurements near the $\beta=0$ threshold limit. Both the $F^{At}_{[d6]}$ and $G^{At}_{[d6]}$ terms scale as ${\cal O}(\beta^2)$ and are strongly suppressed in this limit. The non-vanishing terms for $\beta=0$ are $F^0_{[d6]} $ and $G^1_{[d6]} $. We note that they are proportional to different linear combinations of Wilson coefficients. Denoting for notational simplicity the parameter 
$x_Z = 1/(1-M_Z^2/4m_t^2)$ we find the following functional dependencies in the $\beta=0$ limit:
\bea
F^0_{[d6]} (\beta=0) &\sim & e^2 Q_e Q_t C_{VV}+ 
	 x_Z  g_{vt} \left\{ g_{ve}C_{VV} +g_{ae} C_{VA} \right\} \, , \nonumber \\
G^{1}_{[d6]} (\beta=0) &\sim& e^2 Q_e Q_t C_{VA} + x_Z  g_{vt} \left\{ g_{ve} C_{VA}+g_{ae} C_{VV}\right\} \, .	 
\eea
Both the unpolarized cross section and the spin-correlated terms are proportional to one linear combination of $C_{VV}$ and $C_{VA}$. The single-spin terms depend on a different combination and provide different information on possible new physics. We will see this explicitly later in our numerical results. The contributions of $C_{AA}$ and $C_{AV}$ are suppressed in the $\beta =0$ limit. 

The electroweak dipole operators introduce a new Dirac structure
in the top fermion line, and hence give rise to a spin structure not present in the tree-level SM.
Their contribution to the observables will have the form
\bea
A_{[d6,D]} &=& F^0_{[d6,D]}  +F^1_{[d6,D]} c_t\, , \nonumber
\\C^{(t\bar{t})}_{kk,[d6,D]} &=& F^0_{[d6,D]}c_t^2  +F^1_{[d6,D]} c_t\, ,\nonumber \\
C^{(t\bar{t})}_{rr,[d6,D]} &=&F^0_{[d6,D]} s_t^2 \, , \nonumber \\
C^{(t\bar{t})}_{nn,[d6,D]} &=& 0 \, , \nonumber \\
C^{(t\bar{t})}_{kr,[d6,D]} &=& F^0_{[d6,D]} \frac{2-\beta^2}{2\sqrt{1-\beta^2}}c_t s_t+F^1_{[d6,D]} \frac{s_t}{2\sqrt{1-\beta^2}}c_t \, ,
\eea
where
\bea
F^0_{[d6,D]} &=&\frac{32 \sqrt{2} N_C eQ_e C_{t\gamma} g_{ve} g_{vt}v m_t^3}{D_Z}-8\sqrt{2} N_CQ_tQ_e e^3 C_{t\gamma}v m_t+\frac{32 \sqrt{2} N_C e^2Q_tQ_e C_{tZ} g_{ve}v m_t^3}{ D_Z}\nonumber
\\
&+&\frac{128 \sqrt{2} N_C C_{tZ} g_{vt}v m_t^5 \left(g_{ae}{}^2+g_{ve}{}^2\right)}{D_Z^2}
\, ,\nonumber
\\
F^1_{[d6,D]}&=&\frac{96 \sqrt{2} N_C \beta  e Q_e C_{t\gamma} g_{ae} g_{at}v m_t^3}{3D_Z}+\frac{256 \sqrt{2} N_C\beta  C_{tZ} g_{ae} g_{at} g_{ve}v m_t^5}{D_Z^2} \, .
\eea
For the single-spin observables, we find the following expressions for the non-zero $B_i$:
\bea
B_{k,[d6,D]} &=& G^0_{[d6,D]}c_t  +G^1_{[d6,D]}(1+ c_t^2)\, , \nonumber
\\
B_{r,[d6,D]} &=& -G^0_{[d6,D]}\frac{s_t(2-\beta^2)}{2\sqrt{1-\beta^2}}  +G^1_{[d6,D]} \frac{c_ts_t}{\sqrt{1-\beta^2}}\,,
\eea
where
\bea
G^0_{[d6,D]} &=&\frac{32 \sqrt{2} N_Ce Q_eC_{t\gamma} g_{ae} g_{vt}v  m_t^3}{D_Z}+\frac{32 \sqrt{2} N_C Q_tQ_ee^2 C_{tZ} g_{ae}v m_t^3}{D_Z}+\frac{256 \sqrt{2} N_C C_{tZ} g_{ae} g_{ve} g_{vt}v m_t^5}{D_Z{}^2}\, , \nonumber
\\
G^1_{[d6,D]} &=& \frac{16 \sqrt{2}N_C eQ_e C_{t\gamma}\beta g_{at} g_{ve}v  m_t^3}{D_Z}+\frac{64 \sqrt{2} N_C \beta  C_{tZ}v m_t^5 g_{at}\left(g_{ae}{}^2+g_{ve}{}^2\right)}{D_Z{}^2}\,.
\eea

\section{Top threshold setup}
\label{sec:setup}

In this section, we discuss the details of our analysis of the SMEFT probes possible at a future FCC-ee using inclusive, single-spin, and correlated spin observables. The baseline plans for a future FCC-ee program include approximately 2.65 ab$^{-1}$ of integrated luminosity taken at a center-of-mass energy of $\sqrt{s} = 365$ GeV, as well as 410 fb$^{-1}$ taken at center-of-mass energies between approximately 340 and 345 GeV~\cite{Defranchis:2025auz}. The higher-energy run will allow for an improved determination of the top-quark Yukawa coupling, while the lower-energy runs will focus on the top mass and width. We will consider two scenarios here, consistent with the expectations above:
\begin{itemize}

\item a run of 200 fb$^{-1}$ at the center-of-mass energy $\sqrt{s} = 345$ GeV;

\item a run of 2.65 ab$^{-1}$ at $\sqrt{s} = 365$ GeV.

\end{itemize}
The significantly higher luminosity at $\sqrt{s}=365$ GeV is expected to enhance the sensitivity to new physics, as will be confirmed in our numerical results. For the final state after decay of the top quarks, we include both di-lepton and lepton+jet channels, where the lepton denotes either an electron or a muon. We assume a global reconstruction efficiency of 40\% for the final state, consistent with previous studies~\cite{Maltoni:2024csn}. We use the program {\tt{QQbar\_threshold}}~\cite{Beneke:2016kkb,Beneke:2017rdn} to calculate the total cross-section of $e^+e^-\rightarrow W^+W^-b\bar{b}$ at N$^3$LO QCD plus N$^2$LO electroweak accuracy after initial state radiation. This gives us the following event rates at each energy for the choice $m_t=171.5$ GeV~\cite{Defranchis:2025auz}: 
\bea
\sqrt{s} =345 \, {\rm GeV} &:& 0.533 \, {\rm pb}\, , \nonumber \\
\sqrt{s} =365 \, {\rm GeV} &:& 0.630 \, {\rm pb}\, .
\eea
We take the top-quark and lepton angular distributions from LO and normalize them to these total rates.

It is well known that due to the short lifetime of the top quark, its spin can be determined from the angle of the lepton or jets coming from its decay products. How well the final-state angular distributions reconstruct the spin of the top quark is encoded in a spin-analyzing power. For the lepton+jet final state in the $t\bar{t}$ process, we assume a spin analyzing power of $\alpha_j = 0.64$, while for the di-lepton final state we assume $\alpha_l=1$, consistent with previous studies~\cite{Tweedie_2014}. The numbers of events after reconstruction for each channel and for each center-of-mass energy are collected  
in Table \ref{tab:Ntop}. 
\begin{table}[h!]
\centering
\begin{tabular}{|c|c|c|}
\hline
 & $\sqrt{s}= 345$ GeV &$\sqrt{s}= 365$ GeV\\
 \hline
 Luminosity& $200 \, \text{fb}^{-1}$ & $2.65 \, \text{ab}^{-1}$ \\
 Partonic cross-section& $533\, \text{fb}$ & $630 \, \text{fb}$ \\
    \hline
 Events at parton level&$106\,600$ & $1\,669\,500$ \\
 Reconstructed events ($N_{SM}$)&$42\,640$ & $667\,800$ \\
\hspace{8mm} \rotatebox[origin=c]{180}{$\Lsh$} lepton+jet channel $(N_{SM,lj})$&$6\,139$& $240 \, 387$  \\
 \hspace{4mm} \rotatebox[origin=c]{180}{$\Lsh$} dilepton channel $(N_{SM,ll})$ &$1\,945$& $76 \,170$ \\
 \hline
\end{tabular}
\caption{Estimated event counts for both $\sqrt{s}$ values in the dilepton and lepton+jet channels.}
\label{tab:Ntop}
\end{table}

\subsection{Discussion of observables}

We review here the structure of the differential cross section and show how the coefficients can be determined from a combination of spin and angular measurements. Our approach is motivated by the discussion in~\cite{Bernreuther:2015yna}. We provide a discussion of how a combination of unpolarized, single-spin, and spin-correlated cross sections and asymmetries can best extract different Wilson coefficients, and consider the angular distribution of the top quarks themselves in more detail. Using the definition of the matrix elements in Eq.~(\ref{eq:MEdef}) we can write the differential cross section for the top quark with spin $\hat{s}_t$ and the anti-top spin as $\hat{s}_{\bar{t}}$ as
\be
\frac{1}{\sigma} \frac{d\sigma(\hat{s}_t,\hat{s}_{\bar{t}})}{d c_t} = 
\frac{1}{2}\left(\sigma_A + \hat{s}_t \sigma_{B^{(t)}} + \hat{s}_{\bar{t}} \sigma_{B^{(\bar{t})}} +\hat{s}_t  \hat{s}_{\bar{t}}  \sigma_{C^{(t\bar{t})}}\right)\, .
\ee
The connection between the terms here and those in Eq.~(\ref{eq:MEdef}) is clear from the notation. We consider the following observables that can be formed from a combination of spin measurements and angular distributions.
\begin{enumerate}

\item

In order to see what can be learned from the $t\bar{t}$ angular distributions themselves before 
probing the final-state spin structure, it is convenient to express the differential cross section for $t\bar{t}$ production in terms of Legendre polynomials. We can write
\be
\frac{1}{\sigma}\frac{d\sigma}{d c_t} = \frac{1}{2} \left[ 1+ A_{FB} P_1(c_t) + A_2 P_2 (c_t) \right]\, ,
\ee
where the $P_i$ denote the standard Legendre polynomials and $A_{FB}$ is the forward-backwards asymmetry. The coefficients $A_{FB}$ and $A_2$ can be separated by taking appropriate moments of the differential cross section. Denoting the angular average of a weight function $w(c_t)$ as
\be
 \ww{w(c_t)} = \frac{1}{\sigma} \int_{-1}^1 dc_t \frac{d\sigma}{d c_t}\, ,
\ee
we have
\bea
A_{FB} &=& 3 \ww{P_1(c_t)} \, ,\nonumber \\
A_2 &=& 5 \ww{P_2(c_t)}\, .
\eea
We can refer to Eq.~(\ref{eq:incFdep}) to check which combinations of the $F^i$ parameters the total cross section, $A_{FB}$, and $A_2$ probe. It is straightforward to find 
\bea
\sigma &\propto & 2 \left(1-\frac{\beta^2}{3} \right) F^0 + \frac{4}{3} F^{At},\nonumber \\
A_{FB} &\propto & F^1 \, ,\nonumber \\
A_2 & \propto & \frac{2}{3} \left( \beta^2 F^0 +F^{At} \right)\, .
\eea
We see that the measurement of the forward-backwards asymmetry provides direct access to $F^1$. Due to the $\beta^2$ suppression of $F^{At}$, we expect that inclusive cross section measurements near threshold are primarily sensitive to the $F^0$ coefficient. However, the explicit factor of $\beta^2$ multiplying $F^0$ in the expression for $A_2$ indicates that measurement of this moment of the angular distribution may provide better relative access to the $F^{At}$ term. To estimate the sensitivity of an FCC-ee to new physics using these observables, we include here the statistical uncertainty estimates for each of these observables assuming a total event rate $N$:
\bea
\delta \sigma &\approx & \sqrt{N} \, ,\nonumber \\
\delta A_{FB} &\approx & \sqrt{\frac{3}{N}}\sqrt{1+\frac{2}{5}A_2-\frac{A_{FB}^2}{3}} \, ,\nonumber \\
\delta A_2  &\approx & \sqrt{\frac{5}{N}}\sqrt{1+\frac{2}{7}A_2-\frac{A_2^2}{5}}\, . 
\eea

\item We next consider the single-spin terms where the top-quark spin is measured while the anti-top spin is averaged over. There are two relevant choices of spin vector of the top quark, $\hat{k}$ and $\hat{r}$ (assuming the $\hat{n}$ single-spin coefficient vanishes as discussed earlier). In addition to the usual measurement of the $B^{(t)}$ coefficients obtained from angular distributions of the final-state decay products we also consider these single-spin measurements weighted by the reconstructed top-quark angle. This effectively is a combination of a forward-backward asymmetry with a single-spin measurement. We will see later that these provide additional information for certain Wilson coefficients. We can reconstruct the single-spin coefficients from the single-differential distributions in the final-state angle of either the lepton or jet coming from the top-quark decay~\cite{Bernreuther:2015yna}:
\be
\frac{1}{\sigma} \frac{d \sigma}{dc_s} = \frac{1}{2} \left( 1+\alpha \bar{B}_s c_s \right)
\ee
where $\alpha$ denotes the spin analyzing power discussed previously and the subscript $s$ denotes the angle used to measure the polar angle $c_s$. In our study we use $s=k,r$. The coefficient $\bar{B}_s$ is obtained from $\sigma_{B^{(t)}_{s}}$ after averaging over the top-quark production angle $c_t$. We could get
\be
\bar{B}_s = \frac{3}{\alpha} \langle c_s \rangle.
\ee

Another possibility is to combine measurements of the top-quark kinematics with measurements of the angular distributions of its decay products. In the narrow-width approximation we can express the cross section for the production of a particular final state as the production of the $t\bar{t}$ pair with given spins, multiplied by the differential decay width for the top or anti-top with that spin to decay to the observed final state. We find that the measurements of the double-differential distribution in the top-quark angle combined with the measurement of the final-state decay product angle can lead to useful probes for some combinations of SMEFT coefficients. We can think of these observables as single-spin forward-backward asymmetries. We define them according to
\be
A_{B_s} = \frac{9}{\alpha} \langle c_s c_t \rangle
\ee
where $c_t$ denotes the top-quark polar angle discussed previously.

We include here the statistical uncertainty estimates for each of these observables assuming an event number in the dilepton and lepton+jet channel $N_{l}$, $N_{j}$, respectively~\cite{Tweedie_2014}:
\bea
\delta \bar{B}_{s} &\approx & \frac{1}{\alpha}\sqrt{\frac{3}{N_a}} \, ,\nonumber \\
\delta  A_{B_{s}} &\approx & \frac{1}{\alpha}\frac{3}{\sqrt{N_a}} \, .
\eea
where $a=l,j$.

\item Finally, we consider the case where both the top-quark and anti-top spins are measured. These observables can be reconstructed from appropriate one-dimensional angular distributions as discussed in~\cite{Bernreuther:2015yna}:
\be
\frac{1}{\sigma} \frac{d \sigma}{d\xi_{ij}} = \frac{1}{2} (1- \alpha \bar{C}_{ij} \xi_{ij} ) {\rm ln}\frac{1}{|\xi_{ij}|}
 \ee
where the $\xi_{ij} $ is an appropriate combination of decay angles from the top and anti-top decay products. From this we find
\be
\bar{C}_{ij} = -\frac{9}{\alpha} \langle \xi_{ij} \rangle
\ee
In our study we consider $\bar{C}_{kk},\bar{C}_{rr},\bar{C}_{nn}$ and $\bar{C}_{kr}$. We include here the statistical uncertainty estimates for each of these observables assuming an event number in the dilepton and lepton+jet channel $N_{l}$, $N_{j}$, respectively:
\be
\delta\bar{C}_{ij}  \approx \frac{3}{\alpha_a}\sqrt{\frac{1}{N_a}} \, .
\ee
\end{enumerate}

\section{Numerical results}
\label{sec:numerics}

We begin by turning on a single Wilson coefficient at a time to probe the sensitivity of the various observables considered to potential new physics effects. Our analysis is performed at center-of-mass energies of 345 GeV and 365 GeV, with a single coefficient set to $C_i = 1$. The assumed integrated luminosities are those specified in Table~\ref{tab:Ntop}. In addition to the statistical uncertainties discussed in the previous section we include a 1\% relative systematic uncertainty for all observables in our fits. This choice is an attempt to inject some experimental reality into our fits. We find that the difference between this choice and a statistical-uncertanity only fit is small for all observables except the total event rate.  Our conclusions on the relative strengths of the various observables are unaffected by this choice. In Tables~\ref{tab:1dbound345} and~\ref{tab:1dbound365} we present $95\%$ CL expected bounds on the cutoff scale $\Lambda$ in TeV, truncating the SMEFT expansion of each observable at $\mathcal{O}(1/\Lambda^2)$. The $\mathcal{O}(1/\Lambda^4)$ contributions that come from squaring the dimension-6 effects are found to be negligible for the observables that provide the strongest constraints. We will revisit this issue in more detail in the next section when we consider two-dimensional constraints. 

There are several interesting observations that follow from these tables. We first note that the constraints from $\sqrt{s}=365$ GeV measurements are stronger than their $\sqrt{s}=345$ GeV counterparts. This is not surprising; as we discussed in the previous section this is caused by the significantly higher integrated luminosity anticipated at this energy. We will focus on the higher-energy run in the remainder of our discussion. The strongest expected bounds for each Wilson coefficient  are highlighted in red in the Tables. They range from approximately 4 to 16 TeV, depending on the Wilson coefficient. The weakest bound on the four-fermion Wilson coefficients occurs for $C_{AV}$. This coefficient always enters the $F$ and $G$ structures suppressed by at least one power of $\beta$. Since $\beta \sim 1/3$ at this collider energy this result is not surprising. The strongest bound is for $C_{AA}$ through the measurement of the forward-backward asymmetry. Although this coefficient is also suppressed by $\beta$, it enters the $F^{(1)}$ structure through interference with  the SM photon-exchange, which is numerically much larger than the $Z$-boson contribution. $C_{AV}$ only contributes through interference with the SM $Z$-boson exchange. The expected bound on $C_{VV}$ is strong as well, and this coefficient also enters through interference with the SM photon exchange diagram. The strongest expected bounds are obtained from the total event rate $N$, the forward-backward asymmetry $A_{FB}$, and the single-spin coefficient $\bar{B}_r$. Strong constraints are also expected from the single-spin forward-backward asymmetry $A_{B_k}$. The correlation coefficients $\bar{C}_{ij}$ do not give the strongest probes for any of the Wilson coefficients.

\begin{table}[h!]
\centering
\begin{tabular}{|c|c|c|c|c|c|c|c|c|c|c|c|}
\hline
 &$N$&$A_{FB}$&$A_2$&$\bar{B}_k$&$A_{B_k}$&$\bar{B}_r$&$A_{B_r}$&$\bar{C}_{kk}$&$\bar{C}_{rr}$&${\bar{C}_{nn}}$&${\bar{C}_{kr}}$\\
 \hline
 $\Lambda_{C_{VV}=1}[TeV]$& \color{red}$\mathbf{13}$&$3.5$&$0.44$&$0.89$&$2.2$&$2.4$&$0.48$&$0.18$&$0.24$&$0.16$&$0.69$\\
 \hline
  $\Lambda_{C_{VA}=1}[TeV]$&${\color{red}\mathbf{6.1}}$&$2.1$&$0.21$&$1.1$&$4.7$&$5.0$&$0.60$&$0.088$&$0.12$&$0.076$&$0.41$\\
 \hline
  $\Lambda_{C_{AV}=1}[TeV]$& $0.28$&$\color{red}\mathbf{2.3}$&$0.2$&$1.6$&$0.39$&$0.078$&$0.87$&$0.082$&$0.11$&$0.072$&$0.45$\\
 \hline
  $\Lambda_{C_{AA}=1}[TeV]$&$0.84$&\color{red}$\mathbf{4.8}$&$0.60$&$0.77$&$0.17$&$0.23$&$0.42$&$0.25$&$0.33$&$0.22$&$0.94$\\
 \hline
  $\Lambda_{C_{tZ}=1}[TeV]$&\color{red}$\mathbf{3.3}$&$1.1$&$0.24$&$0.48$&$2.0$&$2.2$&$0.26$&$0.098$&$0.076$&$0.062$&$0.21$\\
 \hline
   $\Lambda_{C_{t\gamma}=1}[TeV]$&\color{red}$\mathbf{7.0}$&$1.9$&$0.50$&$0.50$&$1.2$&$1.3$&$0.27$&$0.21$&$0.16$&$0.13$&$0.38$\\
   \hline 
\end{tabular}
\caption{Expected $95\%$ CL bounds on $\Lambda$ for different $C_i=1$ using $\sqrt{s}=345$ GeV. The strongest expected constraints for each Wilson coefficient choice are highlighted in red.}
\label{tab:1dbound345}
\end{table}

\begin{table}
\centering
\begin{tabular}{|c|c|c|c|c|c|c|c|c|c|c|c|}
\hline
&$N$&$A_{FB}$&$A_2$&$\bar{B}_k$&$A_{B_k}$&$\bar{B}_r$&$A_{B_r}$&$\bar{C}_{kk}$&$\bar{C}_{rr}$&${\bar{C}_{nn}}$&${\bar{C}_{kr}}$\\
 \hline
 $\Lambda_{C_{VV}=1}[TeV]$&\color{red}$\mathbf{14}$& $12$& $2.8$& $3.3$&$4.7$&$4.9$&$1.7$&$1.2$&$1.6$&$1.1$&$2.5$\\
 \hline
  $\Lambda_{C_{VA}=1}[TeV]$&$6.7$& $7.2$& $1.3$&$4.2$&$9.9$&\color{red}$\mathbf{10}$&$2.2$&$0.55$&$0.75$&$0.52$&$1.5$\\
 \hline
  $\Lambda_{C_{AV}=1}[TeV]$&$0.99$& \color{red}$\mathbf{7.9}$&$1.3$&$6.1$&$2.6$&$0.50$&$3.2$&$0.52$&$0.71$&$0.49$&$1.6$\\
 \hline
  $\Lambda_{C_{AA}=1}[TeV]$&$3.0$&\color{red}$\mathbf{16}$& $3.8$&$2.8$&$1.2$&$1.5$&$1.5$&$1.6$&$2.1$&$1.5$&$3.4$\\
 \hline
  $\Lambda_{C_{tZ}=1}[TeV]$&$3.5$& $3.6$&$1.5$&$1.7$&$4.1$&\color{red}$\mathbf{4.4}$&$0.97$&$0.62$&$0.48$&$0.36$&$0.71$\\
 \hline
   $\Lambda_{C_{t\gamma}=1}[TeV]$&\color{red}$\mathbf{7.4}$& $6.5$&$3.2$&$1.8$&$2.6$&$2.5$&$0.9$&$1.3$&$1.0$&$0.85$&$1.3$\\
   \hline 
\end{tabular}
\caption{Expected $95\%$ CL bounds on $\Lambda$ for different $C_i=1$ using $\sqrt{s}=365$ GeV. The strongest expected constraints for each Wilson coefficient choice are highlighted in red.}
\label{tab:1dbound365}
\end{table}

\subsection{Two-coefficient bounds}
\label{subsec:2d}
We now consider fits with two Wilson coefficients turned on simultaneously in order to study correlations in the Wilson-coefficient space, and to find which combinations of observables best resolve possible degeneracies in the parameter space. We again show anticipated $95\%$ CL bounds assuming a $1\%$ systematic uncertainty. Representative plots are shown here, while the complete set of results is provided in the Appendix. In this Section we also study how well the truncation of the SMEFT expansion to the $\mathcal{O}(1/\Lambda^2)$ level works. We focus on the $\sqrt{s}=365$ GeV results here, since as shown in our discussion of single-coefficient bounds they are stronger than the corresponding $\sqrt{s}=345$ GeV results. In these plots we set $\Lambda=1$ TeV and display the limits on the dimensionless Wilson coefficients.

\begin{figure}
 \begin{subfigure}[h]{0.49\textwidth}
     \centering
     \includegraphics[width=\textwidth]{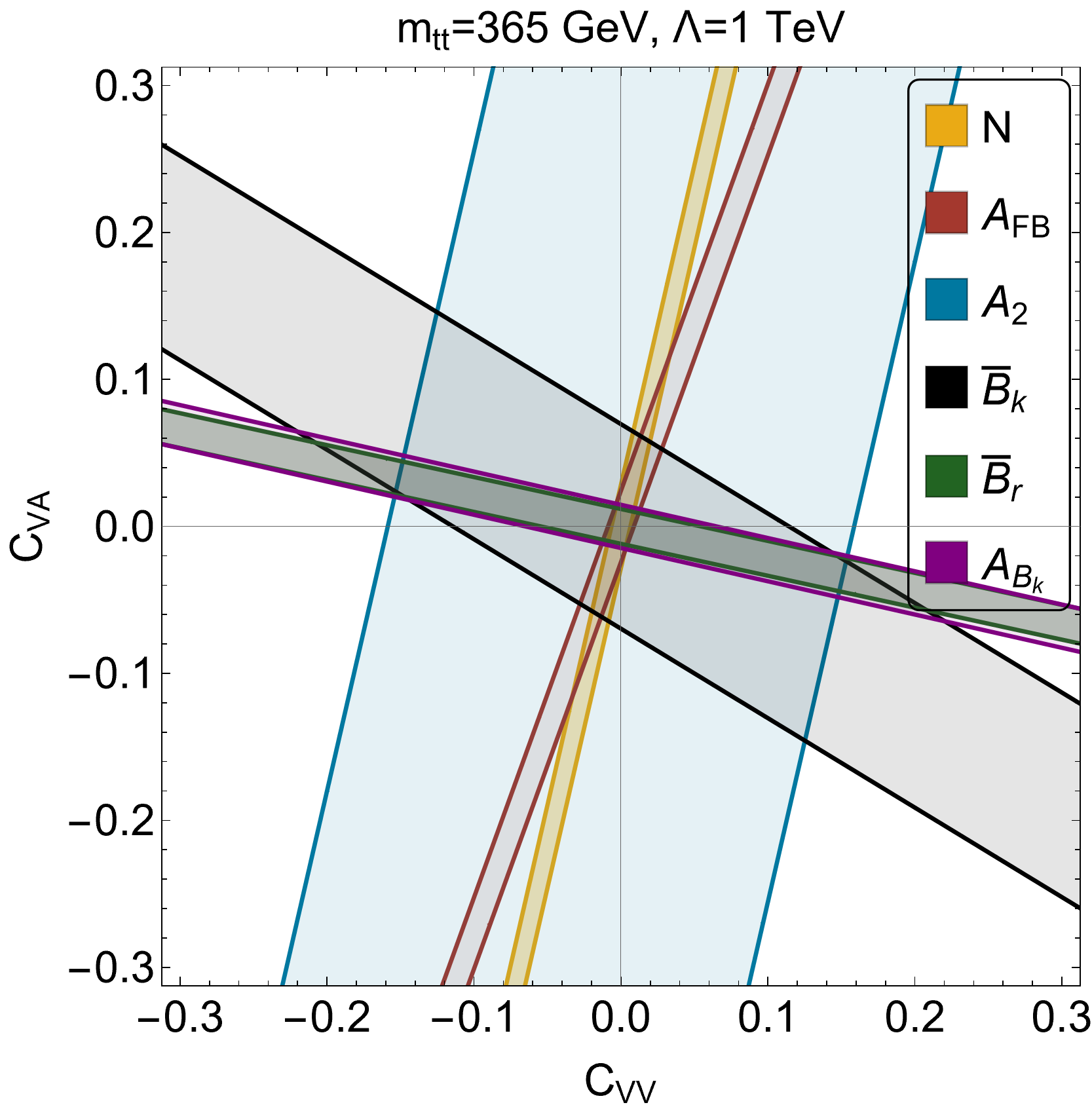}
 \end{subfigure}
 \begin{subfigure}[h]{0.49\textwidth}
     \centering
     \includegraphics[width=\textwidth]{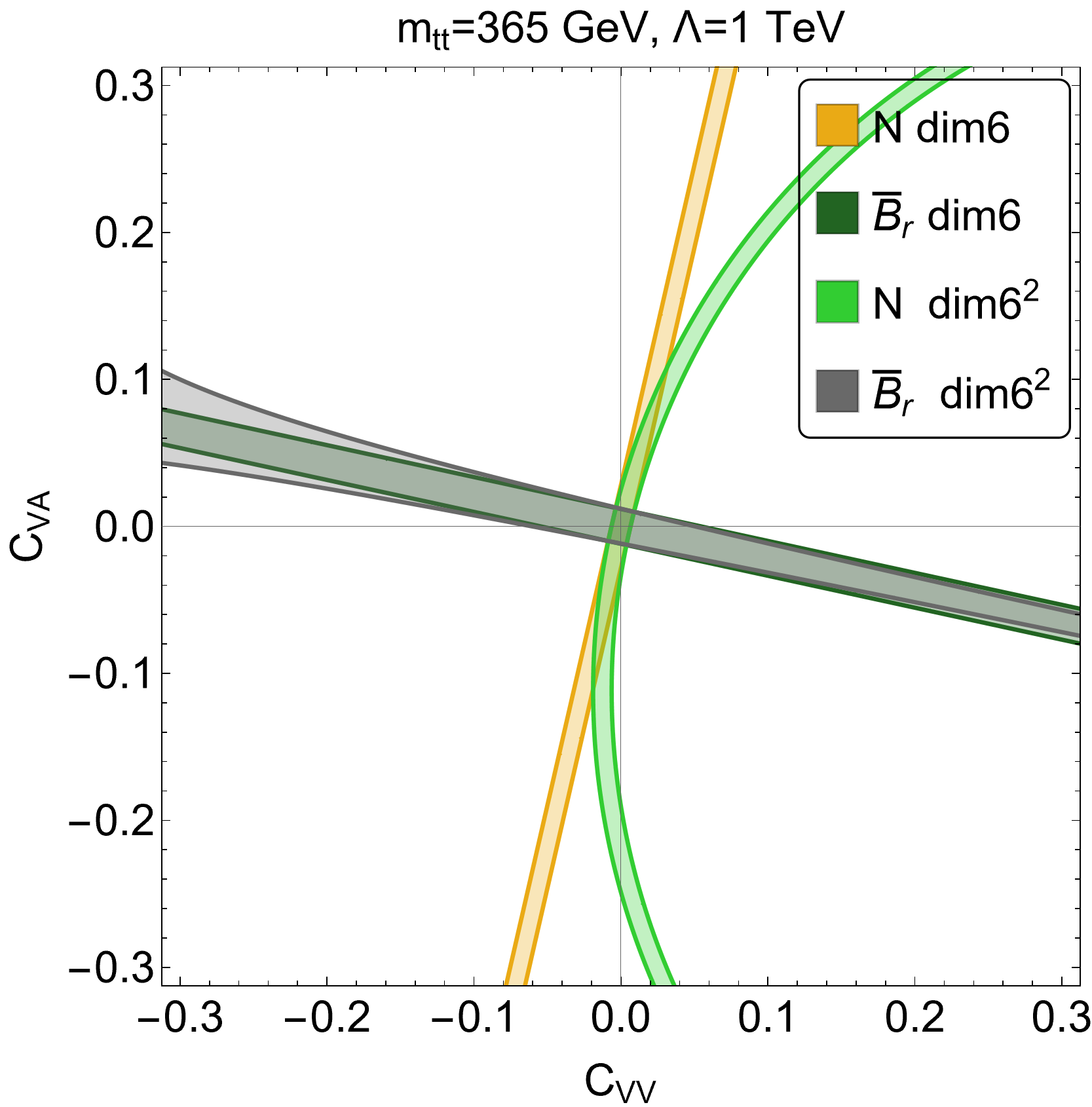}
 \end{subfigure}
 \caption{Regions in the $(C_{VV},C_{VA})$ parameter space expected to be constrained at $\sqrt{s} = 365~\text{GeV}$, based on the observables shown. The allowed regions are shown as shaded areas. The fit in the left panel includes terms up to linear order in $1/\Lambda^2$, while the right panel also accounts for quadratic contributions.}
     \label{fig:2wilson2}
 \end{figure}

We begin by showing the two-dimensional constraints on the choice $C_{VV}$ and $C_{VA}$ provided by several observables in Fig.~\ref{fig:2wilson2} for $\sqrt{s}=365$ GeV. The strongest expected constraints on this parameter space come from the total event rate $N$ and the single-spin coefficient $\bar{B}_r$. Strong constraints also come from $A_{B_k}$ and $A_{FB}$. The various observables probe different linear combinations of the Wilson coefficients as can be seen from the plot. The following pairs of observables show similar correlations, and a choice of one observable from each pair is needed to remove possible flat directions in the parameter space: $(N,A_{FB})$ and $(\bar{B}_r,A_{B_k})$. For these coefficients the potential flat direction can be resolved without spin observables through the measurement of $N$ and $A_{FB}$. The right panel of Fig.~\ref{fig:2wilson2} shows the comparison of the linear dimension-6 fit versus the dimension-6 squared fit. They differ only for larger values of the Wilson coefficients that would be ruled out by a combined fit of the observables. We note that the constraints arising from the $\bar{C}_{ij}$ measurements are significantly weaker than those shown here.

\begin{figure}
\begin{subfigure}[h]{0.49\textwidth}
     \centering
     \includegraphics[width=\textwidth]{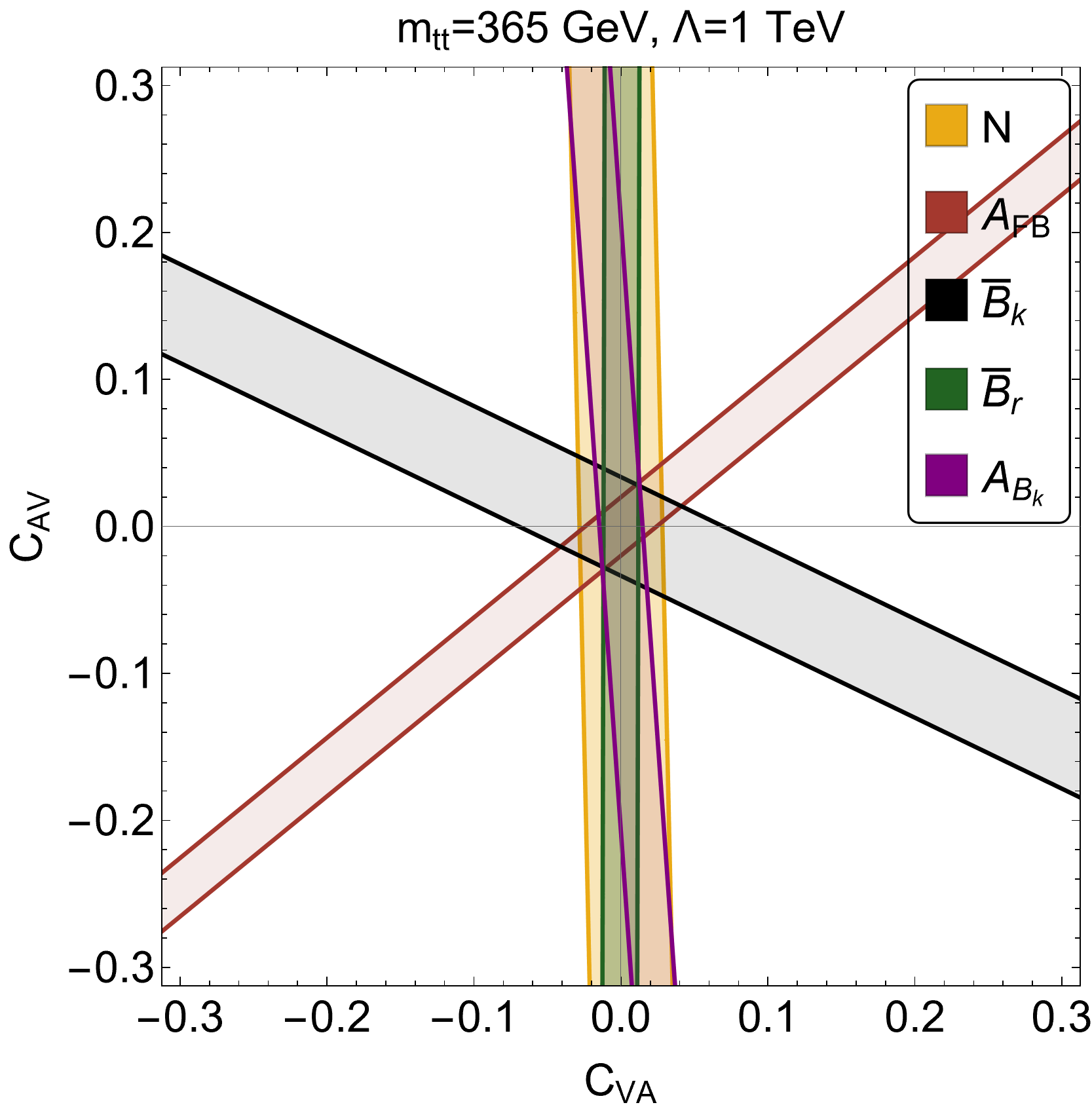}
 \end{subfigure}
 \begin{subfigure}[h]{0.49\textwidth}
     \centering
     \includegraphics[width=\textwidth]{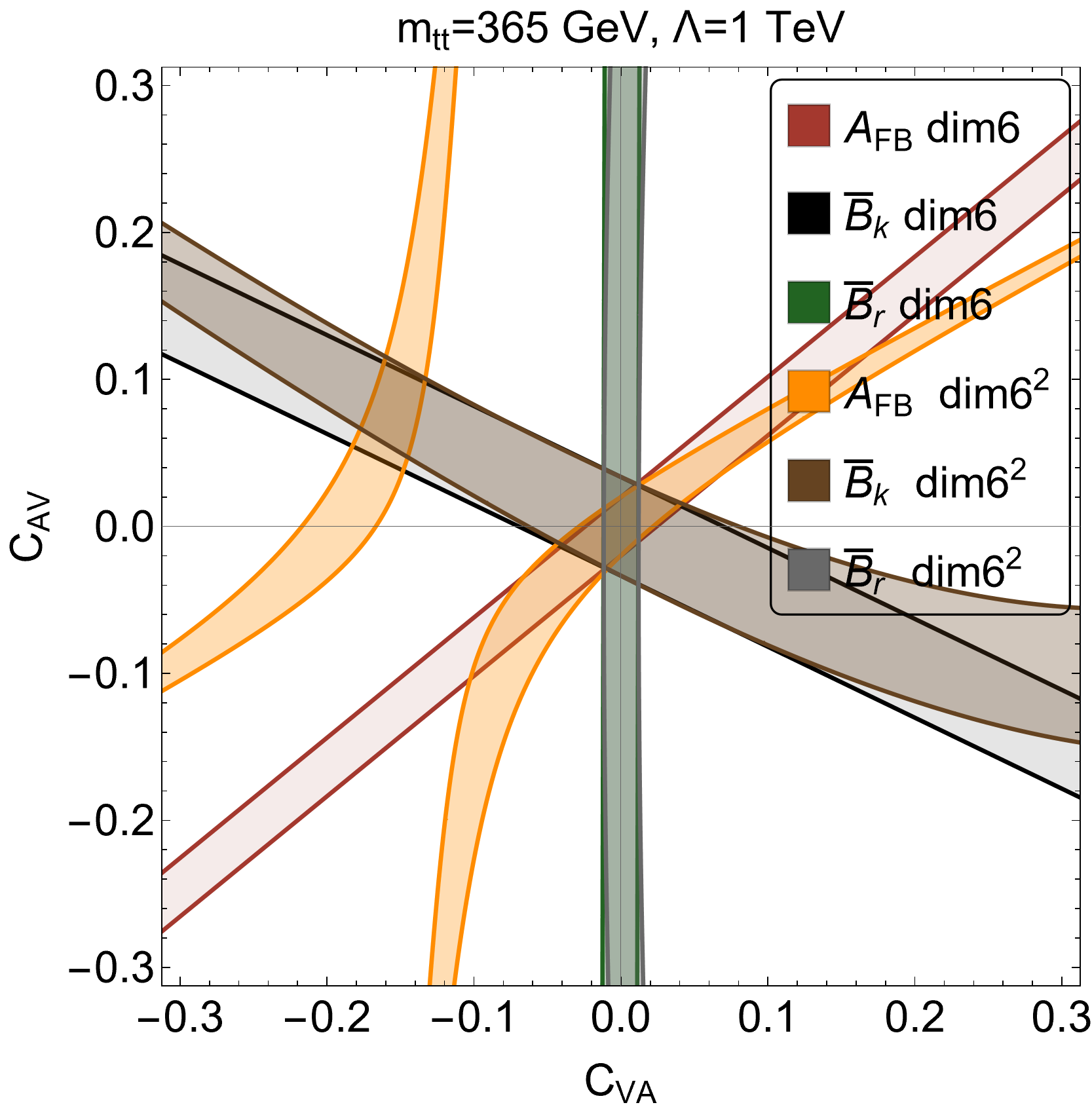}
 \end{subfigure}
      \caption{Regions in the $(C_{VA},C_{AV})$ parameter space expected to be constrained at $\sqrt{s} = 365~\text{GeV}$, based on the observables shown. The allowed regions are shown as shaded areas. The fit in the left panel includes terms up to linear order in $1/\Lambda^2$, while the right panel also accounts for quadratic contributions.}
     \label{fig:2wilson2b}
 \end{figure}

We next turn on the Wilson coefficients $C_{AV}$ and $C_{VA}$ in Fig.~\ref{fig:2wilson2b}. It is interesting to note in this case that $A_{FB}$, $\bar{B}_{r}$ and $\bar{B}_{k}$ all probe different directions in this parameter space. Strong constraints also come from the single-spin forward-backward asymmetry $A_{B_k}$. For this choice of Wilson coefficients, the measurement of at least one of the single-spin coefficients $\bar{B}_{r}$ or $\bar{B}_{k}$ is needed to close all possible flat directions in the parameter space. The comparison of the $1/\Lambda^2$ and $1/\Lambda^4$ fits is shown in the right panel of Fig.~\ref{fig:2wilson2b}. A second allowed region at 95\% CL opens up for the $A_{FB}$ observable for larger values of the Wilson coefficients, coming from the cancellation between linear and quadratic terms in the SMEFT expansion. The measurement of one of the single-spin coefficients can remove this region.

\begin{figure}
  \begin{subfigure}[h]{0.49\textwidth}
     \centering
     \includegraphics[width=\textwidth]{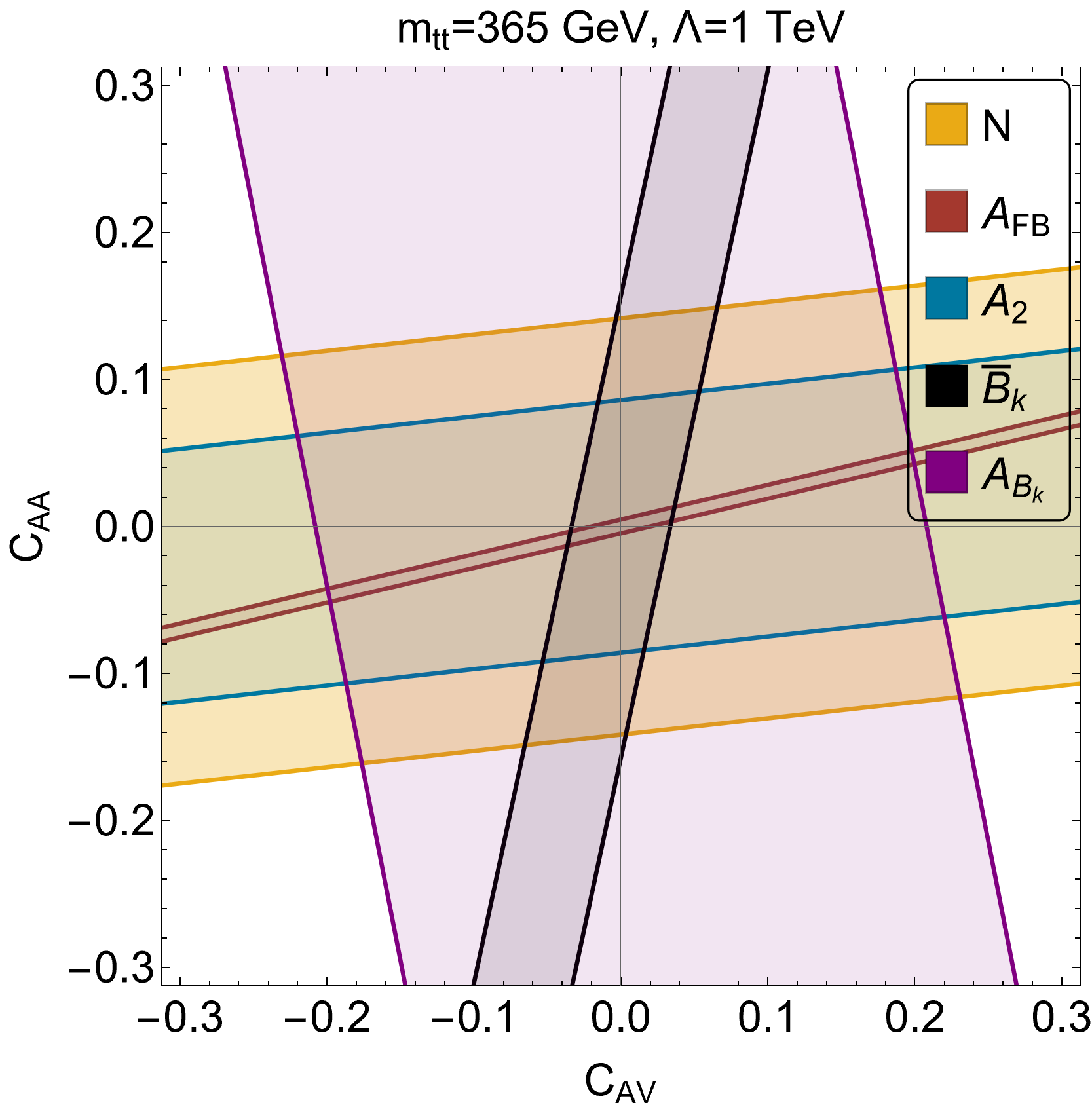}
 \end{subfigure}
 \begin{subfigure}[h]{0.49\textwidth}
     \centering
     \includegraphics[width=\textwidth]{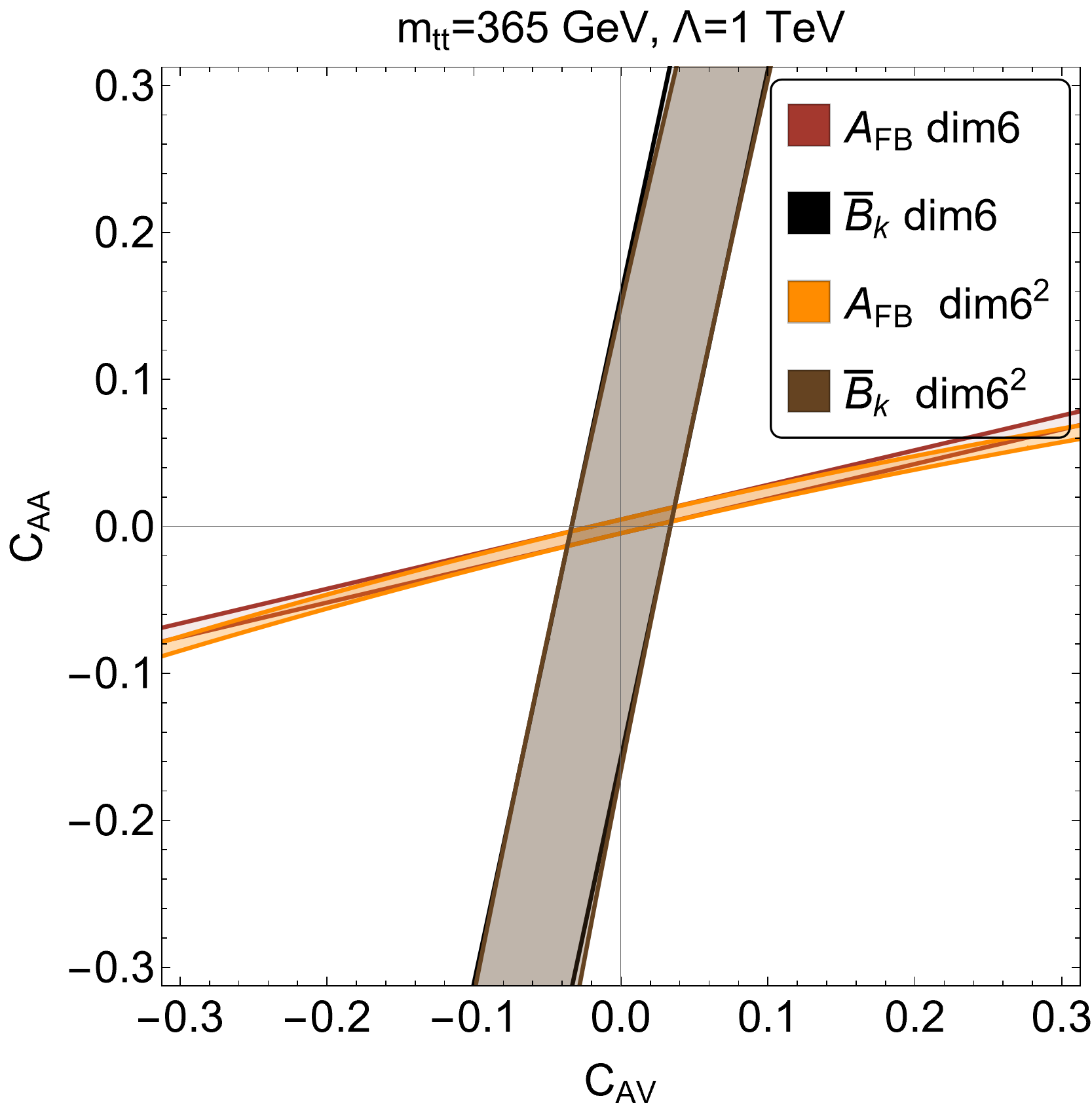}
 \end{subfigure}
      \caption{Regions in the $(C_{AV},C_{AA})$ parameter space expected to be constrained at $\sqrt{s} = 365~\text{GeV}$, based on the observables shown. The allowed regions are shown as shaded areas. The fit in the left panel includes terms up to linear order in $1/\Lambda^2$, while the right panel also accounts for quadratic contributions.}
     \label{fig:2wilson22}
 \end{figure}

We next consider the axial couplings $C_{AA}$ and $C_{AV}$ that come with a $\beta$-suppression in the matrix elements. We recall from the one-dimensional bounds that $C_{AV}$ is the most difficult parameter to probe, and this is clearly seen in Fig.~\ref{fig:2wilson22}. The strongest constraints come from $A_{FB}$ and $\bar{B}_r$, which probe different linear combinations of parameters. $A_{FB}$ essentially probes $C_{AA}$ while $B_{r}$ is most sensitive to $C_{AV}$. While $A_{B_k}$ probes a slightly different linear combination of these two Wilson coefficients, the constraints are weaker than the other two observables. The right panel of Fig.~\ref{fig:2wilson22} shows the $1/\Lambda^2$ and $1/\Lambda^4$ fits. No appreciable difference is seen between these two orders in the SMEFT expansion.

\begin{figure}[H]
 \begin{subfigure}[h]{0.49\textwidth}
     \centering
     \includegraphics[width=\textwidth]{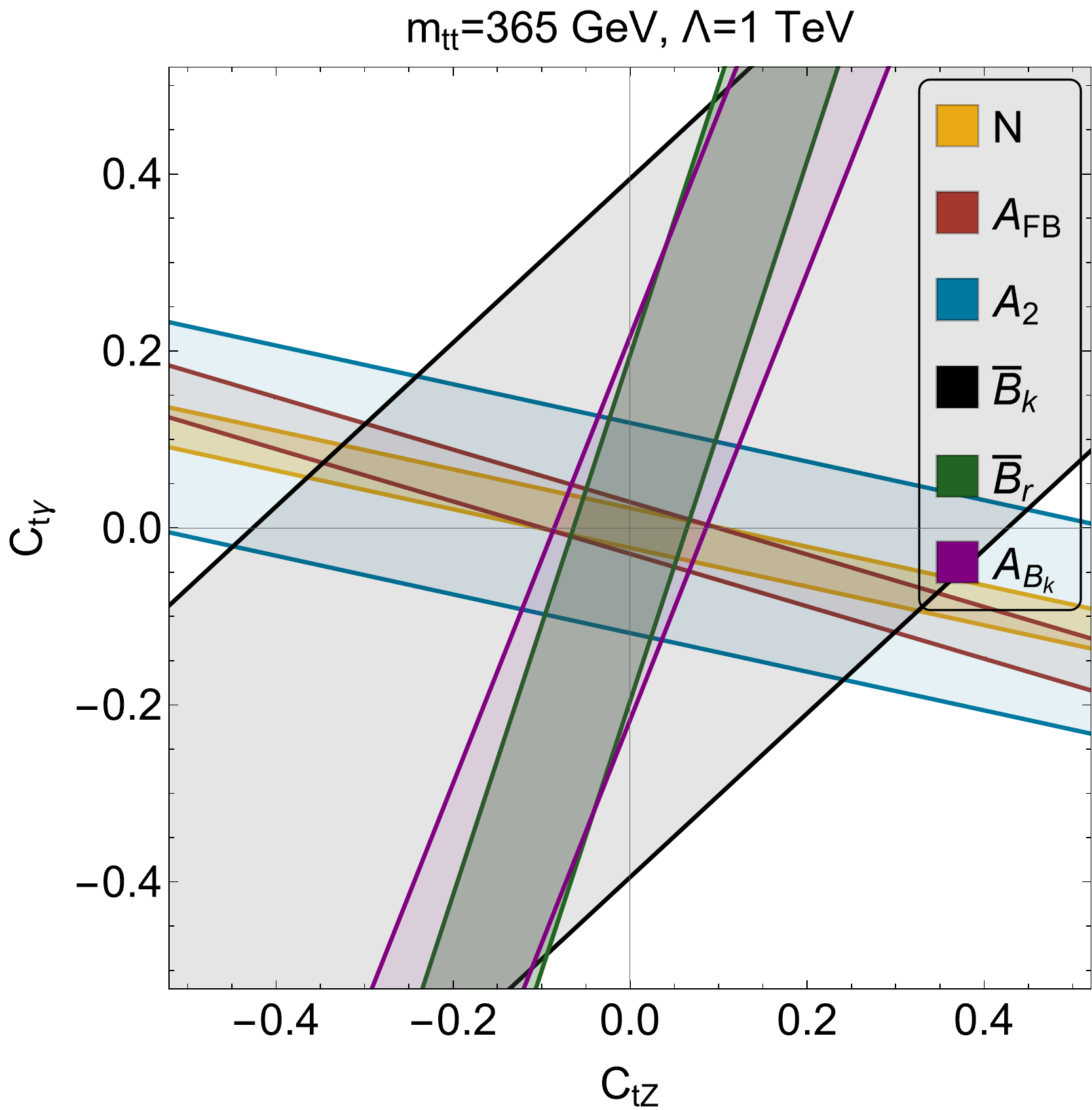}
 \end{subfigure}
  \begin{subfigure}[h]{0.49\textwidth}
     \centering
     \includegraphics[width=\textwidth]{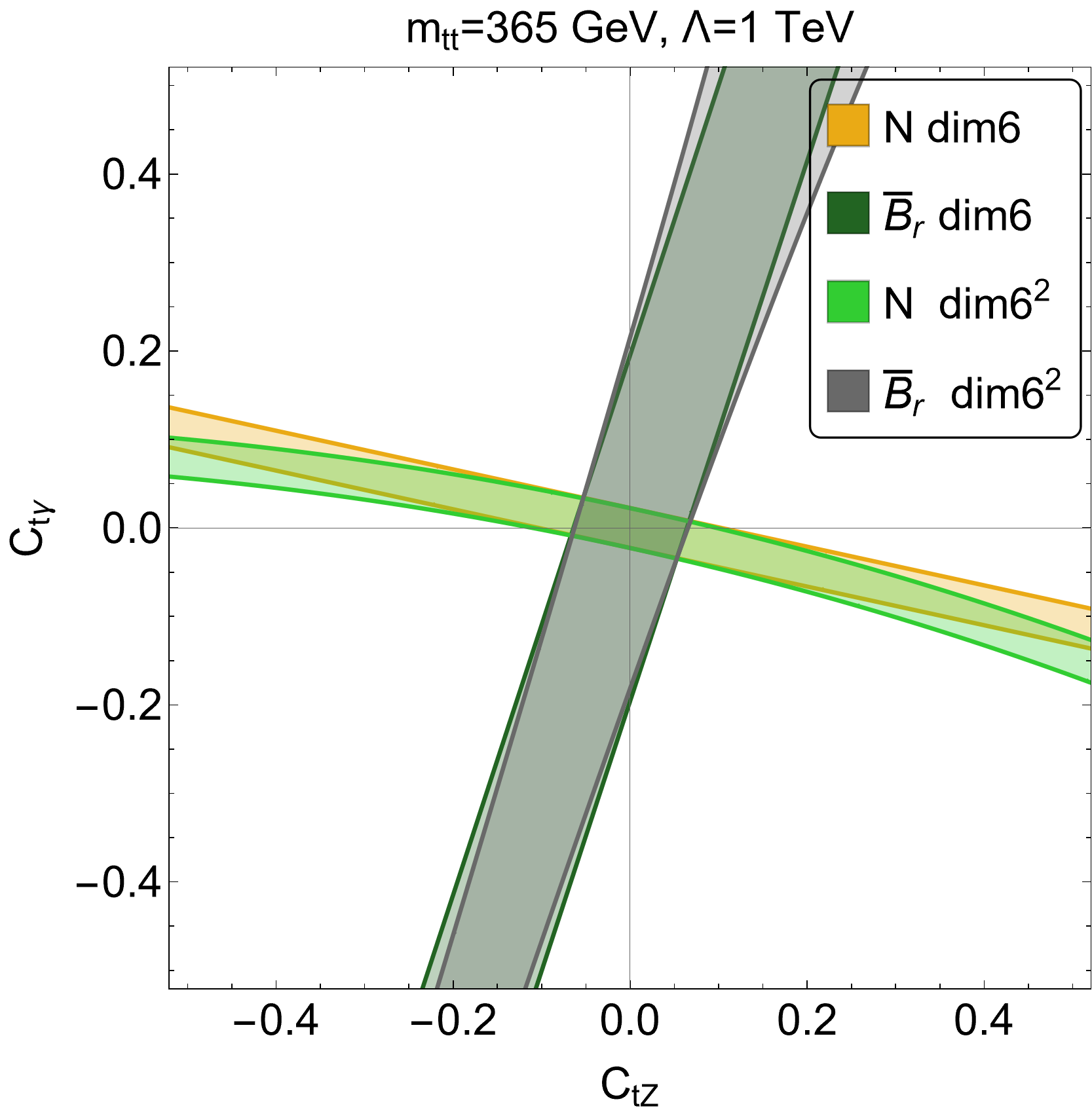}
 \end{subfigure}
             \caption{Regions in the $(C_{tZ},C_{t\gamma})$ parameter space expected to be probed at $\sqrt{s} = 365~\text{GeV}$, based on the observables considered in this study. The allowed regions are shown as shaded areas. The fit in the left panel includes terms up to linear order in $1/\Lambda^2$, while the right panel also accounts for quadratic contributions.}
      \label{fig:2wilson2tztga}
 \end{figure}

We show in Fig.~\ref{fig:2wilson2tztga} the expected constraints in the $(C_{tZ},C_{t\gamma})$ electroweak dipole parameter space. We again observe strong bounds from $N$, $A_{FB}$, $\bar{B}_r$ and $A_{B_k}$. In this case a measurement of either $\bar{B}_r$ or $A_{B_k}$ is needed to close off all directions in parameter space, demonstrating again the need for single-spin measurements. We also note that $A_{B_k}$ provides a small improvement in a narrow region of the $(C_{VA}, C_{tZ})$ plane, while the $A_2$ observable shows useful sensitivity in the $(C_{VV}, C_{t\gamma})$ plane. The right panel of Fig.~\ref{fig:2wilson2tztga} shows that there is no appreciable difference between the dimension-6 and dimension-6 squared fits.

\begin{figure}[H]
     \centering
     \includegraphics[width=0.49\textwidth]{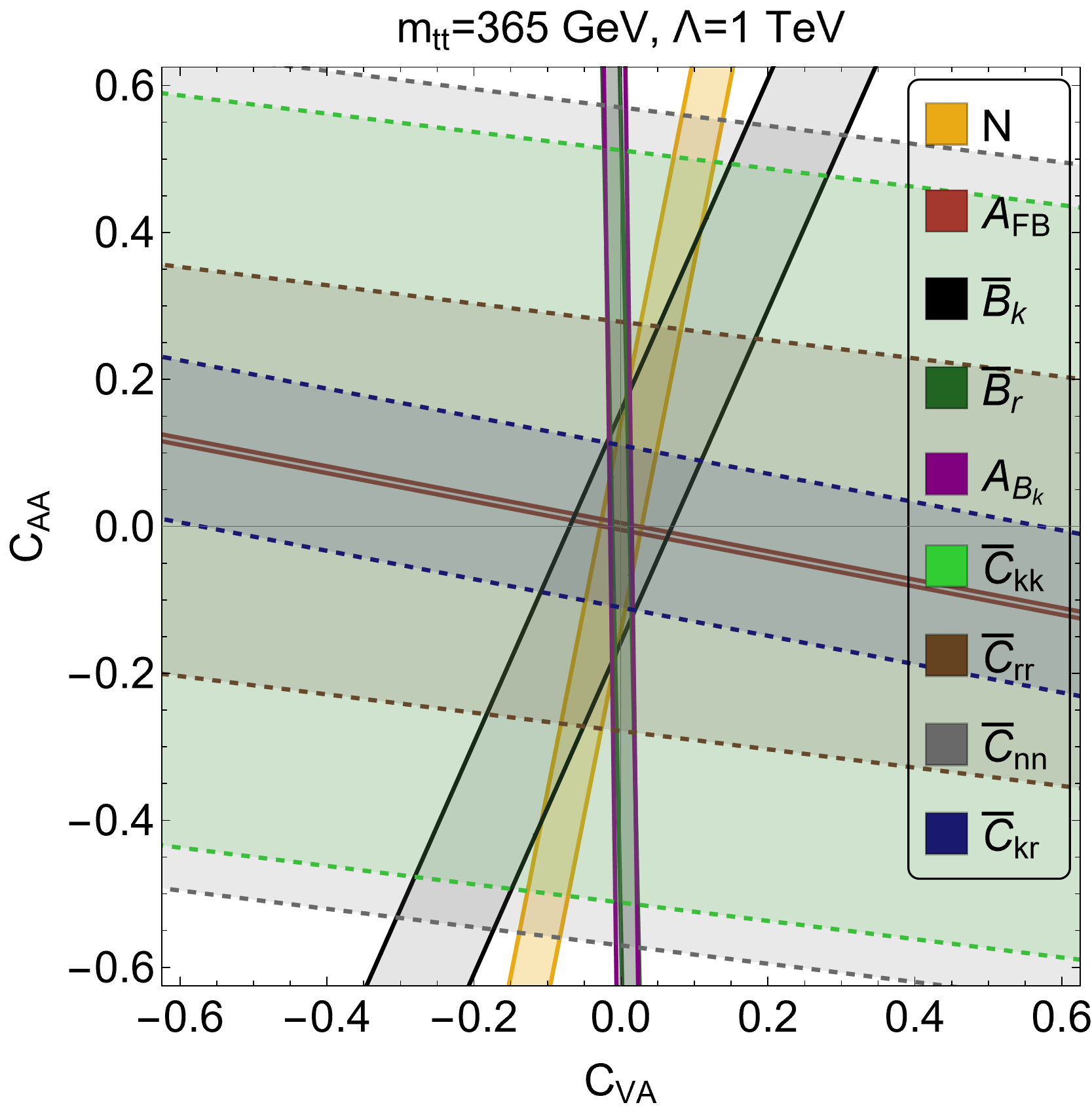}
             \caption{Comparison of bounds from the $\bar{C}_{ij}$ observables (dashed lines) with other observables in the parameter space $(C_{VA},C_{AA})$.}
      \label{fig:Cijcomp}
 \end{figure}

Finally, we compare in Fig.~\ref{fig:Cijcomp} a the expected constraints from the $\bar{C}_{ij}$ coefficients and other observables in the $(C_{VA},C_{AA})$ parameter space. This is representative pair of coefficients; other choices look similar. The expected bounds coming from the $\bar{C}_{ij}$ are weaker than those coming from the other observables, and the $\bar{C}_{ij}$ do not probe different linear combinations of these parameters than the other observables.

\section{Conclusions}
\label{sec:conc}

In this paper we have studied the full spectrum of top-quark observables that can be measured at a future FCC-ee. In comparison with previous work we have focused on single-spin measurements, which we find can provide stronger probes of the heavy new physics parameter space than observables featuring correlated $t\bar{t}$ spins. We show that single-spin measurements probe different combinations of Wilson coefficients in the SMEFT parameter space than other observables, which leads to important complementarity that helps remove flat directions that can appear. For the SMEFT parameter space studied here, a combination of single-spin measurements, the total event rate and the forward-backward asymmetry were sufficient to remove all degeneracies between Wilson coefficients. We also studied the potential of single-spin forward-backward asymmetries and showed that they can  probe parameter regions complementary to other measurements.

We found absolute bounds on the heavy new physics scale exceeding 10 TeV for several of the considered observables. While the explicit numerical values should be taken with caution given the lack of a detailed experimental study, our results suggest that single-spin measurements will be a powerful probe of new physics in the top sector at an FCC-ee. It would be interesting to extend this analysis to SMEFT operators that appear first at loop level that may be difficult to access directly elsewhere, to see what spin measurements can teach us.

\section*{Acknowledgements}

We thank Haiqiang Tao for valuable discussions. H.~C. and F.~P. are supported by the U.S. Department of Energy,
Office of High Energy Physics, under contract No. DE-SC0010143. H.~C. is also supported by the Center for Frontiers in Nuclear Science through a joint postdoctoral fellowship. This research was supported in part through the computational resources and staff contributions provided
for the Quest high performance computing facility at Northwestern University which is jointly supported by the Office
of the Provost, the Office for Research, and Northwestern University Information Technology. 
 
\bibliographystyle{h-physrev}
\bibliography{Tspin}
\appendix
\section{Common factors for spin correlation coefficients}
In this Appendix, we present the expressions for the common factors $F^{[i]}$ that enter the coefficients $A$ and $C_{ij}$, and also give the factors $G^{[i]}$ appearing in the coefficients $B_i$.
   
\bea
F^0_{[SM]} &=& N_C e^4  Q_t^2 
+\frac{8 e^2 g_{ve} g_{vt} N_C Q_tQ_e m_t^2}{D_Z}  
+\frac{16 g_{vt}^2 N_C \left(g_{ae}^2+g_{ve}^2\right) m_t^4}{D_Z^2}\, ,\nonumber
\\
F^1_{[SM]} &=& \frac{16  N_C g_{ae}g_{at} m_t^2\beta}{D_Z}\left(\frac{8 g_{ve} g_{vt} m_t^2}{D_Z}+   N Q_tQ_te^2  \right)\, ,\nonumber
\\
F^{At}_{[SM]} &=&\frac{16 \beta ^2 g_{at}^2 N_C \left(g_{ae}^2+g_{ve}^2\right) m_t^4}{D_Z^2}\, ,
\nonumber
\\
G^0_{[SM]} &=& 8N_C \frac{m_t^2}{D_Z}\beta\left[4g_{vt} g_{at}(g_{ve}^2+ g_{ae}^2)+e^2Q_tQ_eg_{ve} g_{at}\right]\, ,\nonumber
\\
G^1_{[SM]} &=& \frac{16N_C g_{ae} g_{vt} m_t^2}{D_Z}\left[4g_{ve} g_{at}\frac{m_t^2}{D_Z} +e^2 Q_tQ_e  \right]\, ,\nonumber
\\
G^{At}_{[SM]} &=&\frac{64 N_C \beta ^2 g_{ae} g_{at}^2 g_{ve} m_t^4}{D_Z^2}\, .
\eea
We have introduced the abbreviation $D_Z = D_Z = 4 m_t^2 -M_Z^2(1-\beta^2)$ for the $Z$-boson propagator contribution.
For SMEFT top-pair production, the $1/\Lambda^2$ terms are given in the main text, while the $1/\Lambda^4$ results are provided here. The common factors  $F^{[i]}$ at this order are given by
\bea
F^0_{[d8]} &=& \frac{16 N_C m_t^4 \left(C_{VA}{}^2+C_{VV}{}^2\right)}{\left(\beta ^2-1\right)^2}+\frac{16N_C e C_{HV} v^2 m_t^4 (g_{ae} C_{VA}+g_{ve} C_{VV})}{\left(\beta ^2-1\right) c_ws_w D_Z }\nonumber
\\
&+&\frac{4N_C e^2 C_{HV}{}^2 v^4 \left(g_{ae}^2+g_{ve}^2\right) m_t^4}{c_w^2 s_w^2 D_Z^2}
\, ,\nonumber
\\
F^1_{[d8]} &=& \frac{64 N_C \beta m_t^4 ( C_{AA} C_{VV}+  C_{AV} C_{VA})}{\left(\beta ^2-1\right)^2}+
\frac{32N_C \beta e^2  g_{ae} g_{ve} C_{HA} C_{HV} v^4 m_t^4}{c_w^2 s_w^2 D_Z^2}
\nonumber
\\
&+&\frac{32N_C \beta  e v^2 m_t^4 (g_{ae} C_{AV} C_{HV}+g_{ae} C_{HA} C_{VV}+g_{ve} C_{AA} C_{HV}+g_{ve} C_{HA} C_{VA})}{\left(\beta ^2-1\right) c_ws_w D_Z }\, ,\nonumber
\\
F^{At}_{[d8]} &=&\frac{16 N_C m_t^4 \beta ^2 \left(C_{AA}{}^2+C_{AV}{}^2\right)}{\left(\beta ^2-1\right)^2}+\frac{16 N_C \beta ^2 e v^2 m_t^4 C_{HA} (g_{ae} C_{AA}+g_{ve} C_{AV})}{\left(\beta ^2-1\right) c_w s_w D_Z}
\nonumber
\\
&+&\frac{4N_C \beta ^2 e C_{HA}{}^2 v^4 \left(g_{ae}^2+g_{ve}^2\right) m_t^4}{c_w^2 s_w^2 D_Z^2}\, .
\eea
When the contribution of the electroweak dipole operator is included, the square of this operator induces additional contributions to 
$A$, which take the form
\bea
A_{[d8,D]}=F^0_{[d8,D]}+F^1_{[d8,D]}c_t+F^2_{[d8,DD]}c_t^2
\eea
\bea
F^0_{[d8,D]} &=& \frac{64 \sqrt{2} C_{tZ}N_Cv m_t^5 \left[\left(1-\beta ^2\right) e C_{HV} v^2 \left(g_{ae}^2+g_{ve}^2\right)-2 c_w D_Z s_w (g_{ae} C_{VA}+g_{ve} C_{VV})\right]}{\left(\beta ^2-1\right) c_w D_Z^2 s_w}\nonumber
\\
&+&\frac{16 \sqrt{2}N_C C_{t\gamma}e v  m_t^3 \left[2 c_w D_Z C_{VV} s_w+\left(\beta ^2-1\right) e g_{ve} C_{HV} v^2\right]}{\left(\beta ^2-1\right) c_w D_Z s_w}\nonumber
\\
&+&\frac{128N_C  \left(\beta ^2-2\right) C_{tZ}{}^2 v^2 \left(g_{ae}^2+g_{ve}^2\right) m_t^6}{\left(\beta ^2-1\right) D_Z^2}+\frac{8 N_C \left(\beta ^2-2\right) e^2 C_{t\gamma }{}^2 v^2  m_t^2}{\beta ^2-1}\nonumber
\\
&-&\frac{64N_C \left(\beta ^2-2\right) e g_{ve} C_{t\gamma } C_{tZ} v^2  m_t^4}{\left(\beta ^2-1\right) D_Z}
\, ,\nonumber
\\
F^1_{[d8,D]} &=& \frac{16N_C C_{t\gamma}\sqrt{2} \beta  e v  m_t^3 \left(2 c_w D_Z C_{AA} s_w+\left(\beta ^2-1\right) g_{ae} e C_{HA} v^2\right)}{\left(\beta ^2-1\right) c_w D_Z s_w}
\nonumber
\\
&-&\frac{128 \sqrt{2} N_CC_{tZ}\beta  v m_t^5 \left[c_w D_Z g_{ae} C_{AV} s_w+c_w D_Z g_{ve} C_{AA} s_w+\left(\beta ^2-1\right) g_{ae} e g_{ve} C_{HA} v^2\right]}{\left(\beta ^2-1\right) c_w D_Z^2 s_w}\, ,
\nonumber
\\
F^2_{[d8,DD]} &=& \frac{128N_C \beta ^2 C_{tZ}{}^2 v^2 \left(g_{ae}^2+g_{ve}^2\right) m_t^6}{\left(\beta ^2-1\right) D_Z^2}+\frac{8N_C \beta ^2 e^2 C_{t\gamma }{}^2 v^2  m_t^2}{\beta ^2-1}+\frac{64N_C \beta ^2 e g_{ve} C_{t\gamma } C_{tZ} v^2  m_t^4}{D_Z(1-\beta^2)} \nonumber \\
\eea
The common factors  $G^{[i]}$ at order $1/\Lambda^4$ are given by
\bea
G^0_{[d8]} &=&\frac{32 N_C \beta  m_t^4 (C_{AA} C_{VA}+C_{AV} C_{VV})}{\left(\beta ^2-1\right)^2}+\frac{8N_C \beta  e^2 C_{HA} C_{HV} v^4 \left(g_{ae}^2+g_{ve}^2\right) m_t^4}{c_w^2 D_Z^2 s_w^2}\nonumber
\\
&+&
\frac{16N_C \beta  e v^2 m_t^4 (g_{ae} C_{AA} C_{HV}+g_{ae} C_{HA} C_{VA}+g_{ve} C_{AV} C_{HV}+g_{ve} C_{HA} C_{VV})}{\left(\beta ^2-1\right) c_w D_Z s_w}
\, ,\nonumber
\\
G^1_{[d8]} &=&\frac{64 N_C C_{VA} C_{VV}  m_t^4}{\left(1-\beta ^2\right)^{2}}+\frac{32N_C e v^2 m_t^4 C_{HV} (g_{ae} C_{VV}+g_{ve} C_{VA})}{\left(\beta ^2-1\right) c_w D_Z s_w}+\frac{16N_C g_{ae} e^2 g_{ve} v^4 m_t^4C_{HV}^2}{c_w^2 D_Z^2 s_w^2}\, ,\nonumber
\\
G^{At}_{[d8]} &=&\frac{64 N_C m_t^4 \beta ^2 C_{AA} C_{AV}}{\left(\beta ^2-1\right)^2}+\frac{32N_C e v^2 m_t^4 \beta ^2 C_{HA} (g_{ae} C_{AV}+g_{ve} C_{AA})}{\left(\beta ^2-1\right) c_w D_Z s_w}+\frac{16N_C \beta ^2 g_{ae} e^2C_{HA}^2 g_{ve} v^4 m_t^4}{c_w^2 D_Z^2 s_w^2}\, ,
\nonumber
\\
G^0_{[d8,D]} &=& -\frac{128 \sqrt{2}N_C C_{tZ} v m_t^5 \left[c_wD_Zg_{ae} C_{VV} s_w+c_wD_Zg_{ve} C_{VA} s_w+\left(\beta ^2-1\right) g_{ae} e g_{ve} C_{HV} v^2\right]}{\left(\beta ^2-1\right) c_w D_Z^2 s_w}\nonumber
\\
&+&\frac{16 \sqrt{2} N_Ce C_{t\gamma} v  m_t^3 \left(2 c_wD_ZC_{VA} s_w+\left(\beta ^2-1\right) g_{ae} e C_{HV} v^2\right)}{\left(\beta ^2-1\right) c_wD_Zs_w}-\frac{128N_C g_{ae} e C_{t\gamma} C_{tZ} v^2  m_t^4}{D_Z}\nonumber
\\
&+&
\frac{512N_C g_{ae} g_{ve} C_{tZ}^2 v^2 m_t^6}{D_Z^2}
\, ,\nonumber
\\
G^1_{[d8,D]} &=& \frac{8 \sqrt{2}N_C C_{t\gamma}\beta  e v  m_t^3 \left[2 c_wD_ZC_{AV} s_w+\left(\beta ^2-1\right) e g_{ve} C_{HA} v^2\right]}{\left(\beta ^2-1\right) c_wD_Zs_w}+
\nonumber
\\
&+&\frac{32 \sqrt{2}N_C C_{tZ}\beta  v m_t^5 \left[\left(1-\beta ^2\right) e C_{HA} v^2 \left(g_{ae}^2+g_{ve}^2\right)-2 c_wD_Zs_w (g_{ae} C_{AA}+g_{ve} C_{AV})\right]}{\left(\beta ^2-1\right) c_w D_Z^2 s_w}\, .
\eea
\section{Plots}
\begin{figure}[t]
     \centering
 \begin{subfigure}{0.43\textwidth}
     \centering
     \includegraphics[width=\textwidth]{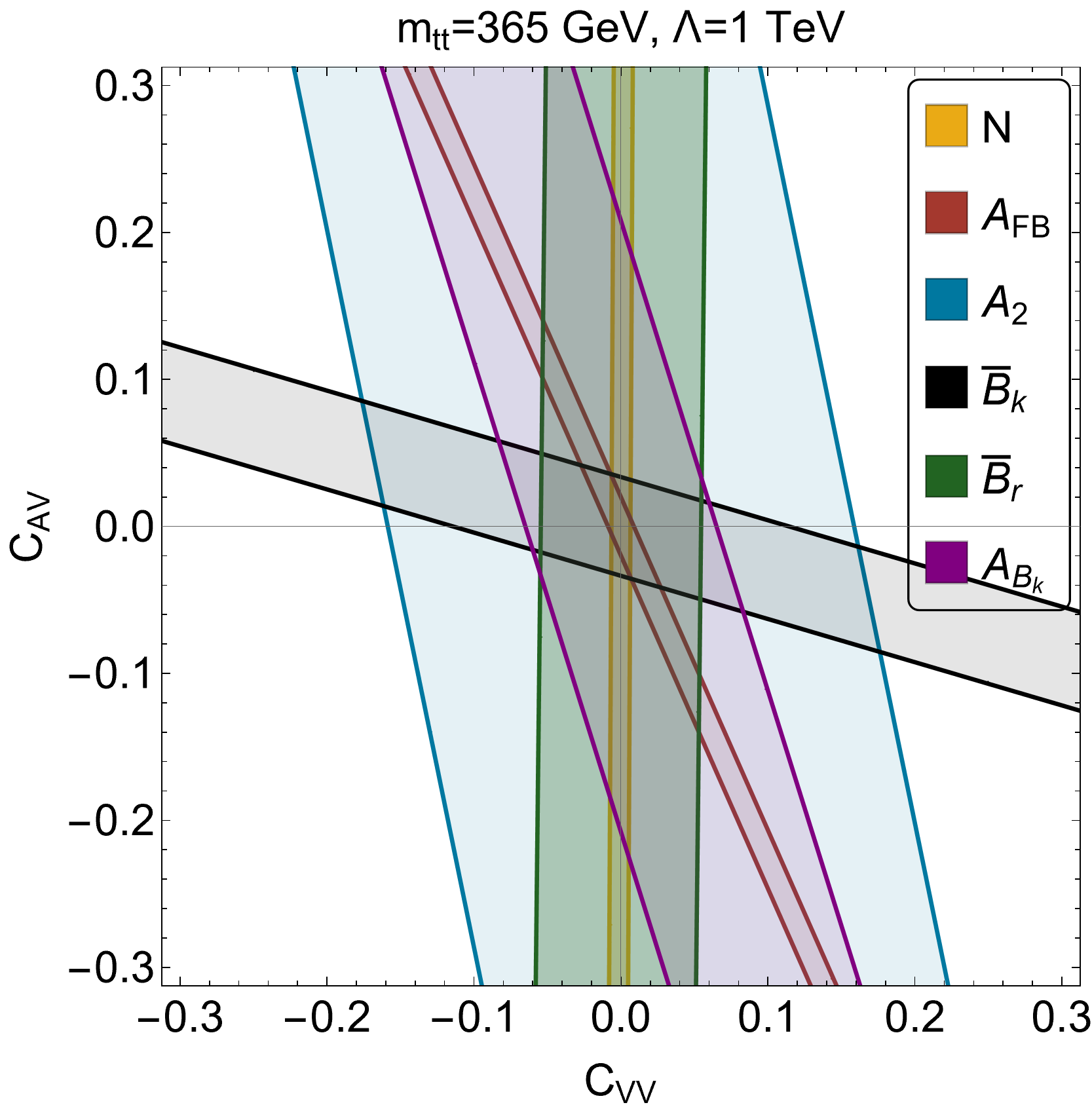}
 \end{subfigure}
  \begin{subfigure}{0.43\textwidth}
     \centering
     \includegraphics[width=\textwidth]{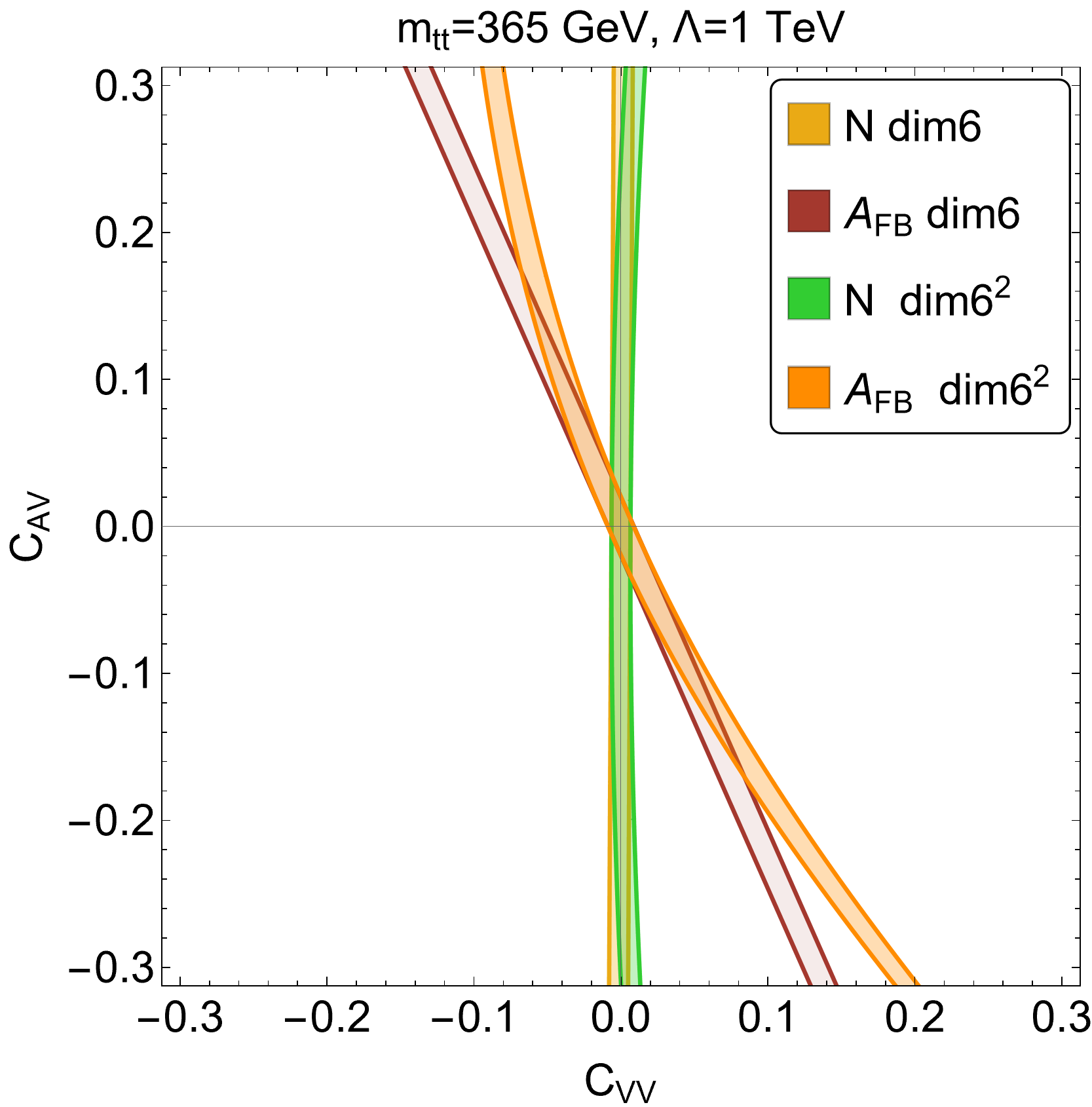}
 \end{subfigure}
  \begin{subfigure}{0.43\textwidth}
     \centering
     \includegraphics[width=\textwidth]{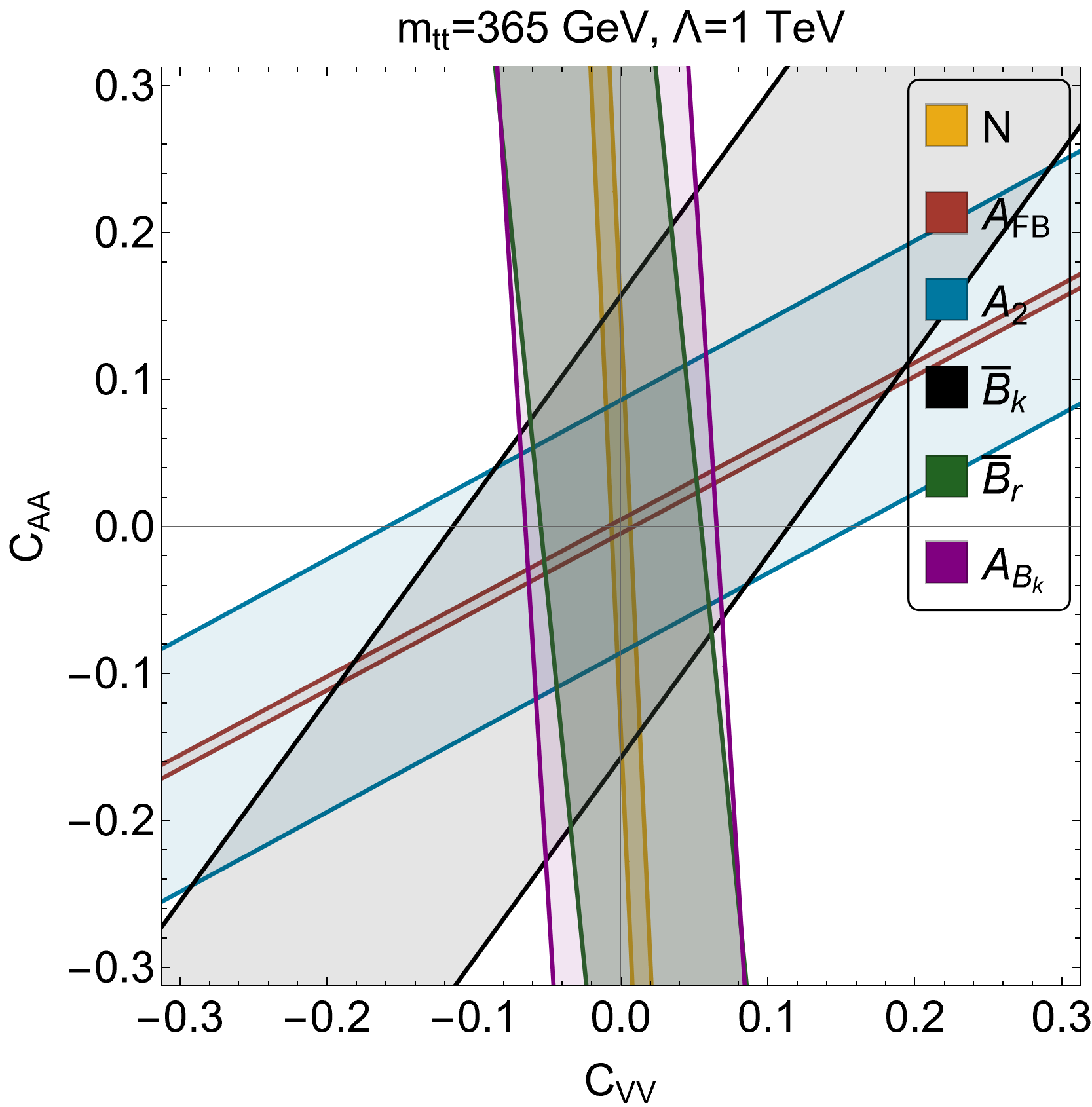}
 \end{subfigure}
  \begin{subfigure}{0.43\textwidth}
     \centering
     \includegraphics[width=\textwidth]{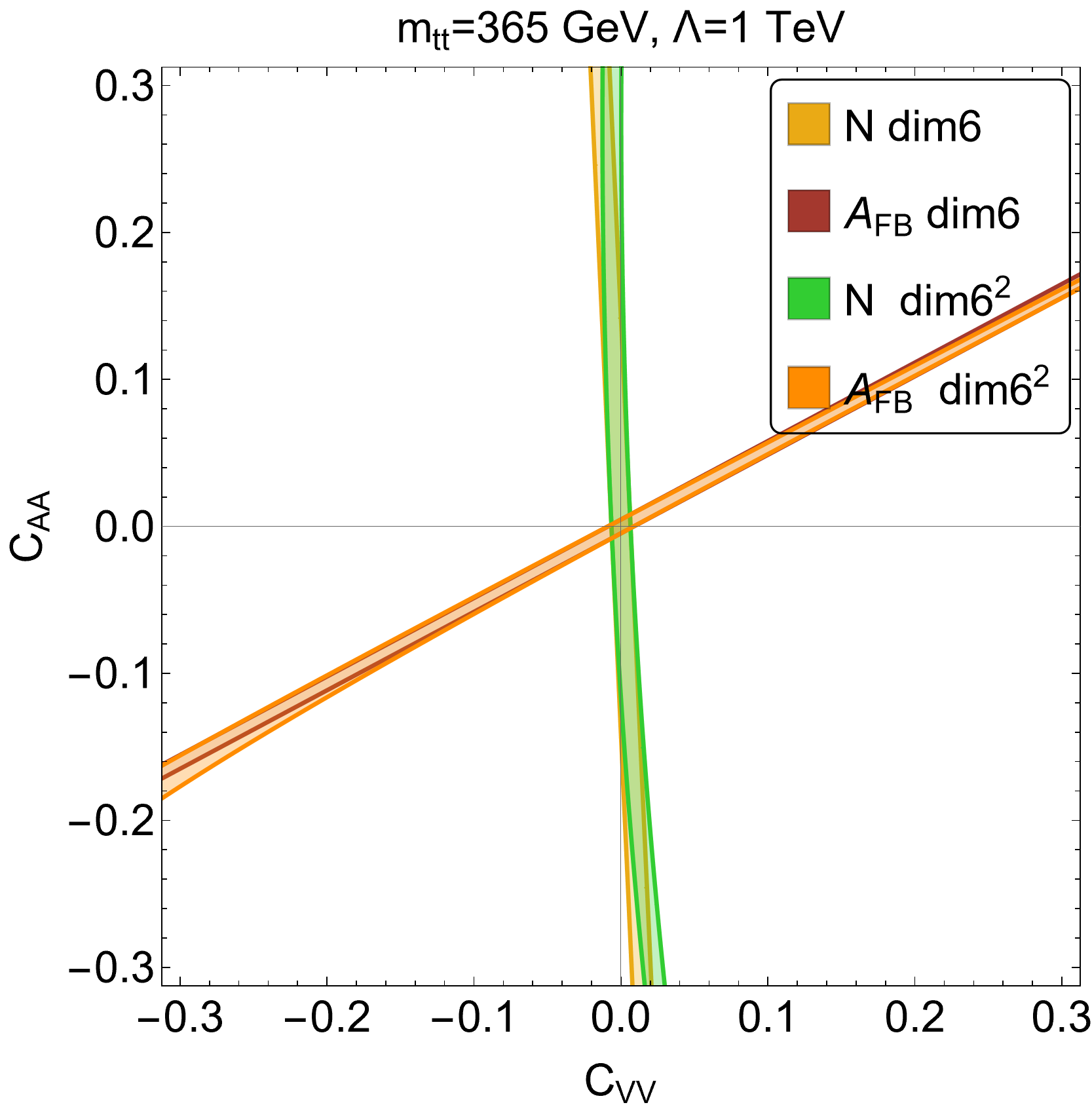}
 \end{subfigure}
    \begin{subfigure}{0.43\textwidth}
     \centering
     \includegraphics[width=\textwidth]{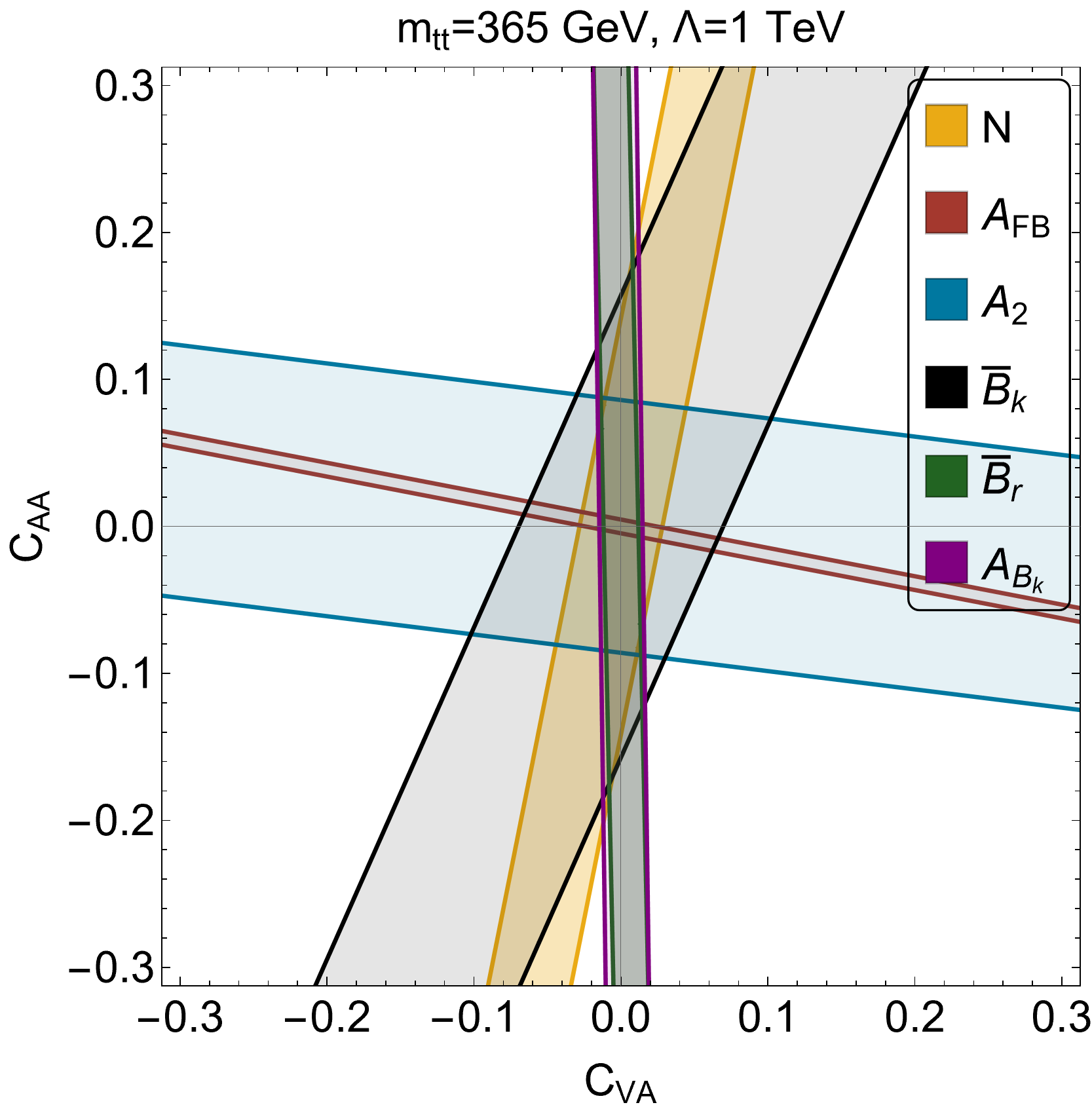}
 \end{subfigure}
 \begin{subfigure}{0.43\textwidth}
     \centering
     \includegraphics[width=\textwidth]{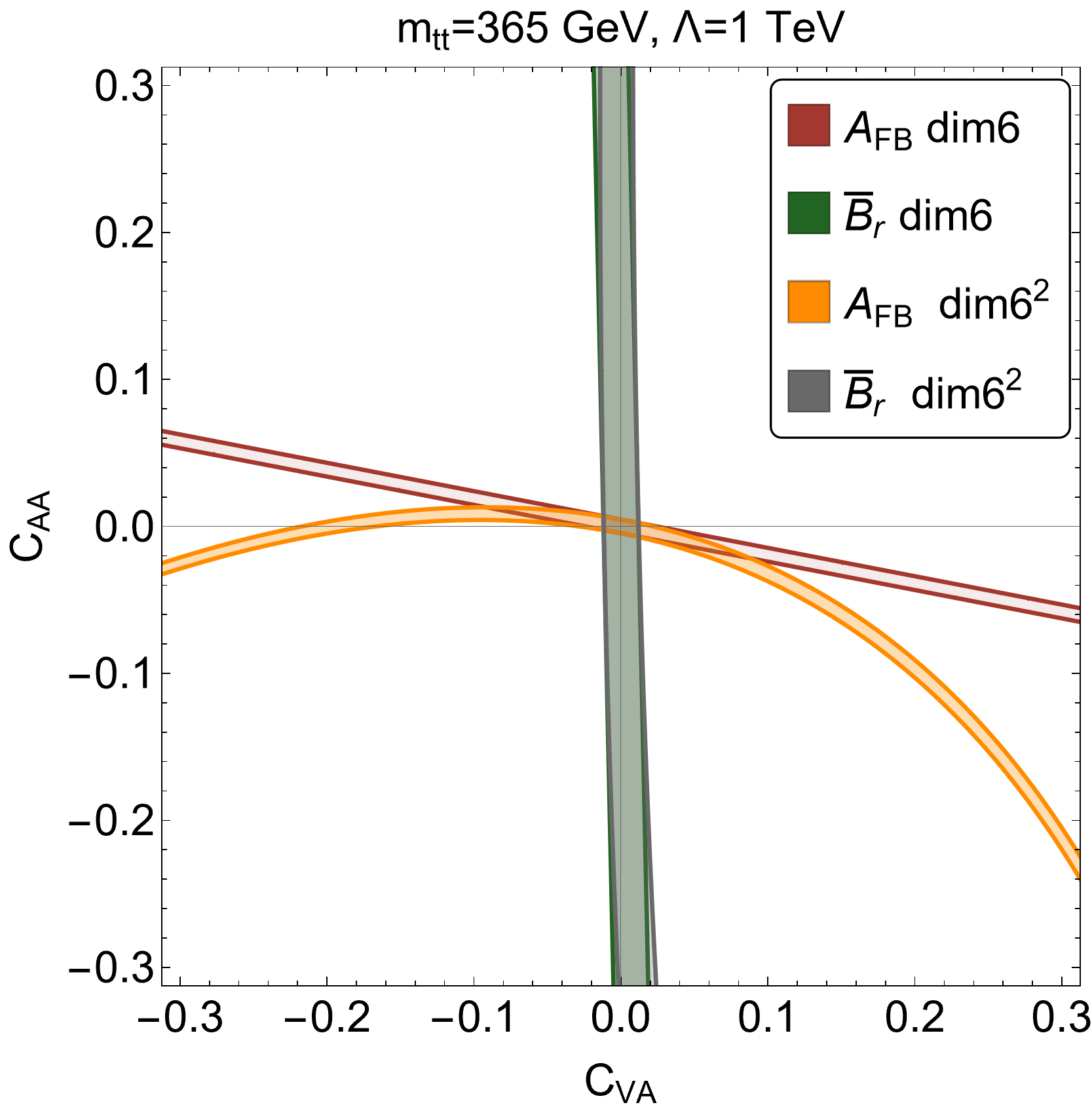}
 \end{subfigure}
     \caption{Same as in the Fig.~\ref{fig:2wilson2} in Section~\ref{subsec:2d} of the main text, for the pairs $(C_{VV}, C_{AV})$, 
    $(C_{VV}, C_{AA})$, and $(C_{VA}, C_{AA})$.}
     \label{fig:2wilson3}
 \end{figure}

\begin{figure}[t]
     \centering
 \begin{subfigure}{0.43\textwidth}
     \centering
     \includegraphics[width=\textwidth]{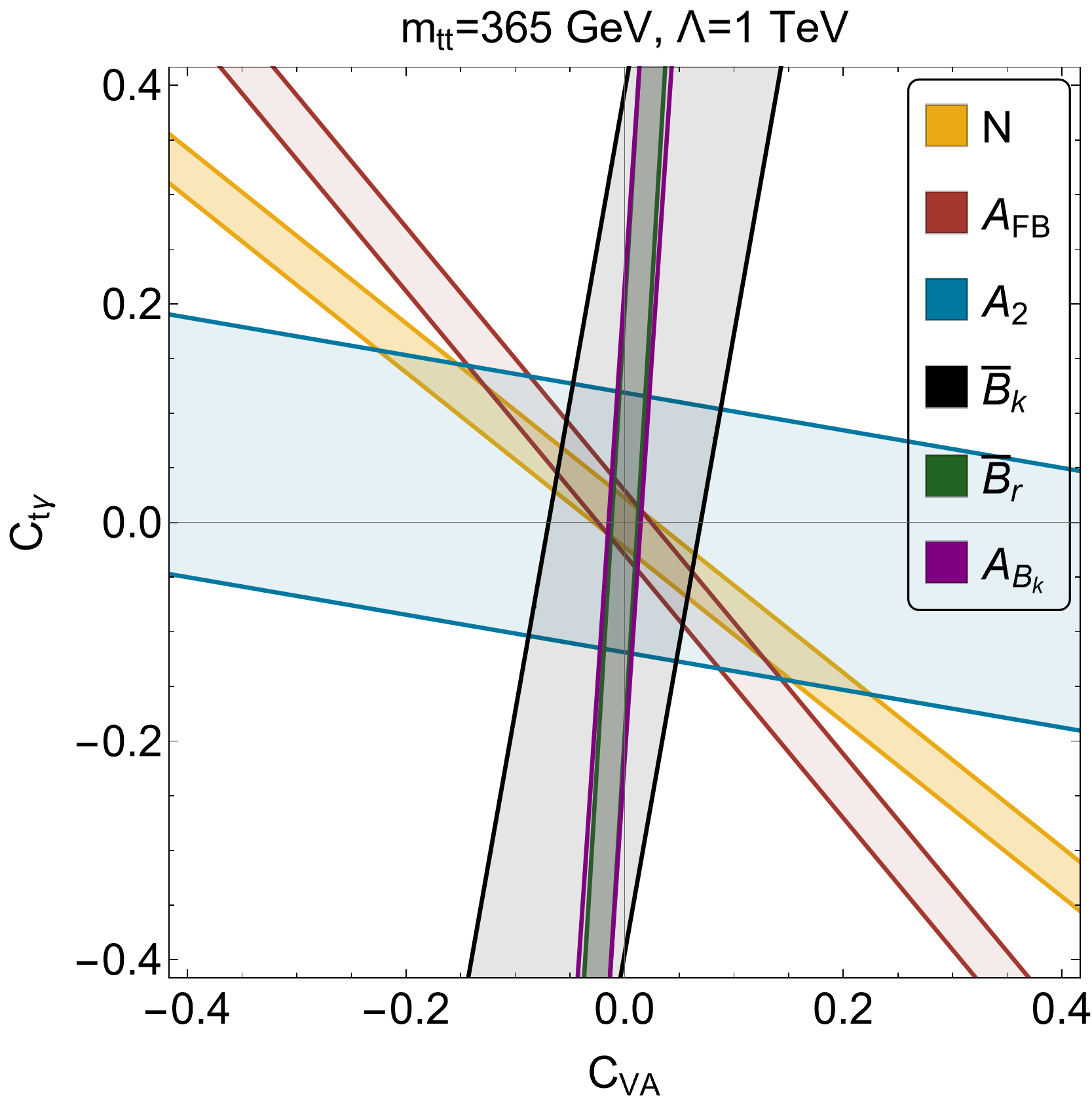}
 \end{subfigure}
  \begin{subfigure}{0.43\textwidth}
     \centering
     \includegraphics[width=\textwidth]{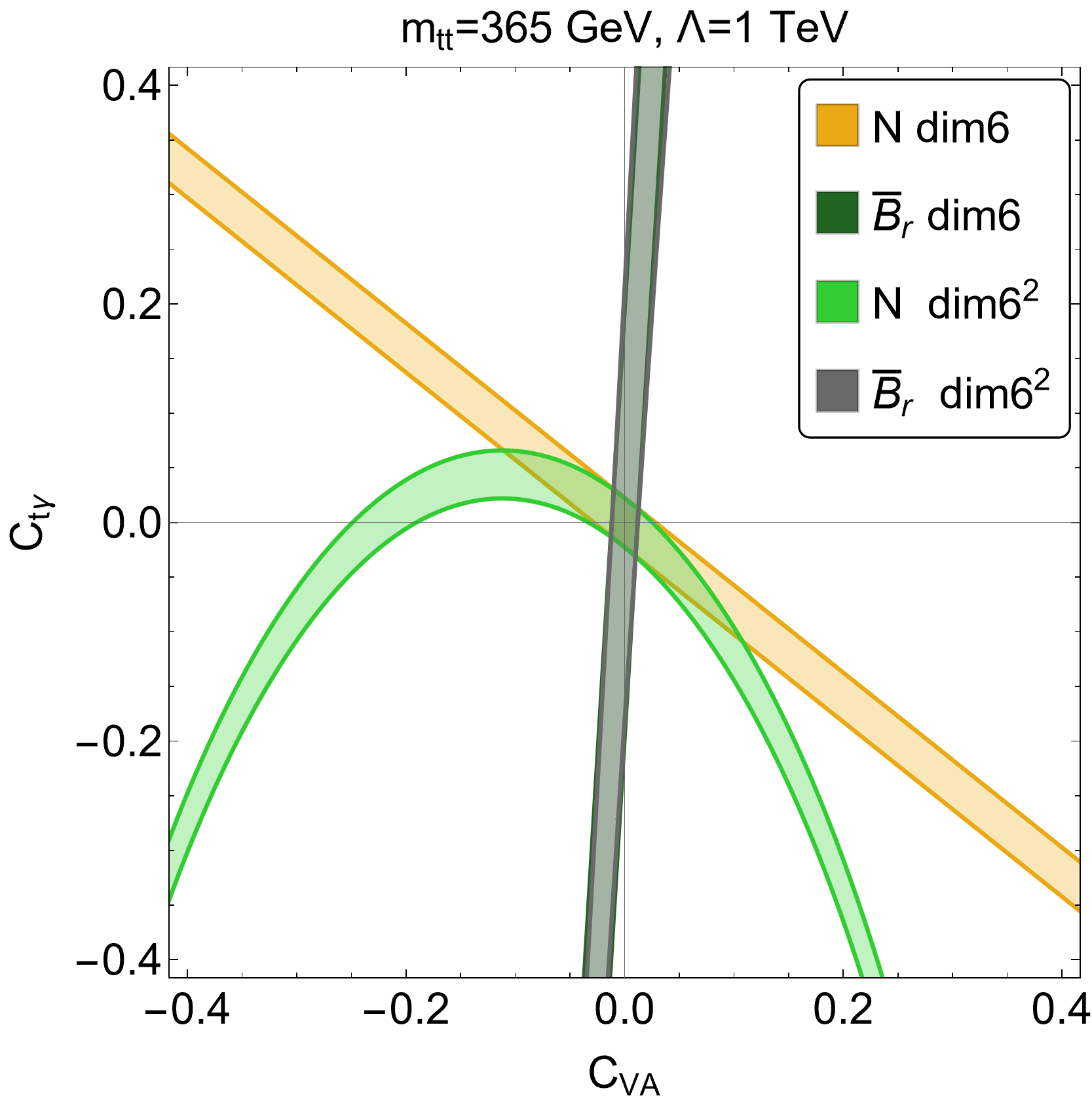}
 \end{subfigure}
  \begin{subfigure}{0.43\textwidth}
     \centering
     \includegraphics[width=\textwidth]{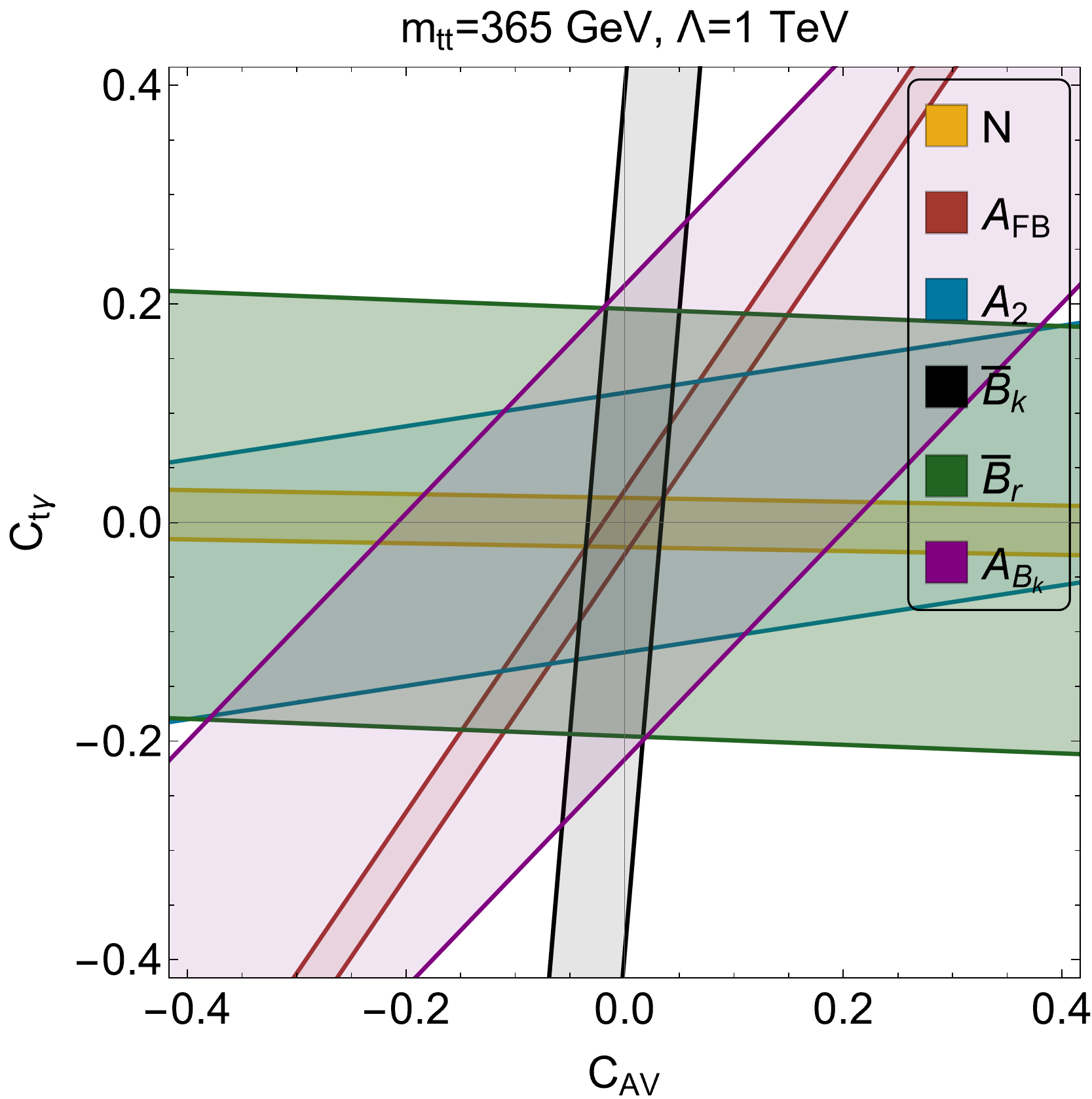}
 \end{subfigure}
  \begin{subfigure}{0.43\textwidth}
     \centering
     \includegraphics[width=\textwidth]{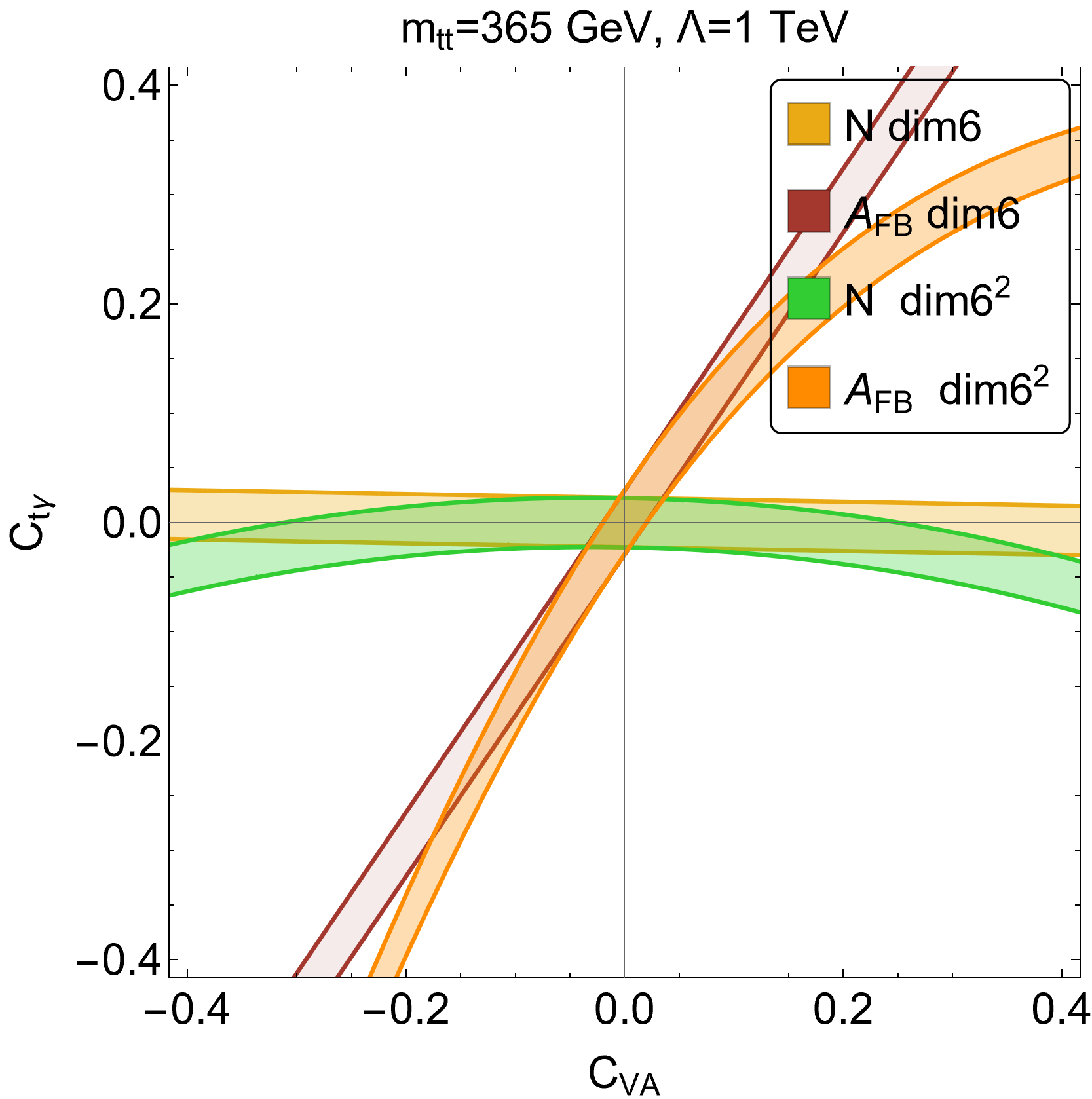}
 \end{subfigure}
    \begin{subfigure}{0.43\textwidth}
     \centering
     \includegraphics[width=\textwidth]{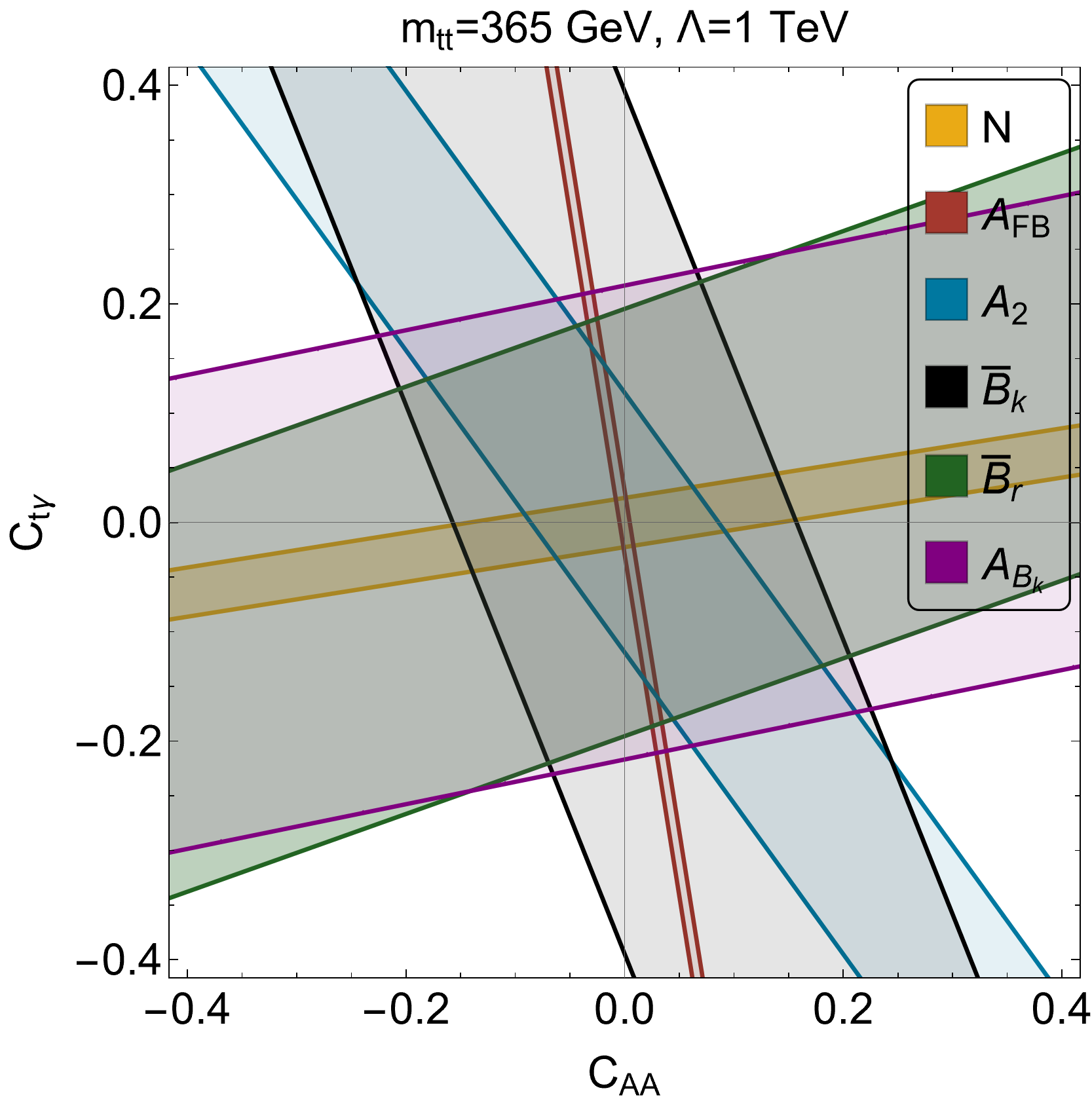}
 \end{subfigure}
 \begin{subfigure}{0.43\textwidth}
     \centering
     \includegraphics[width=\textwidth]{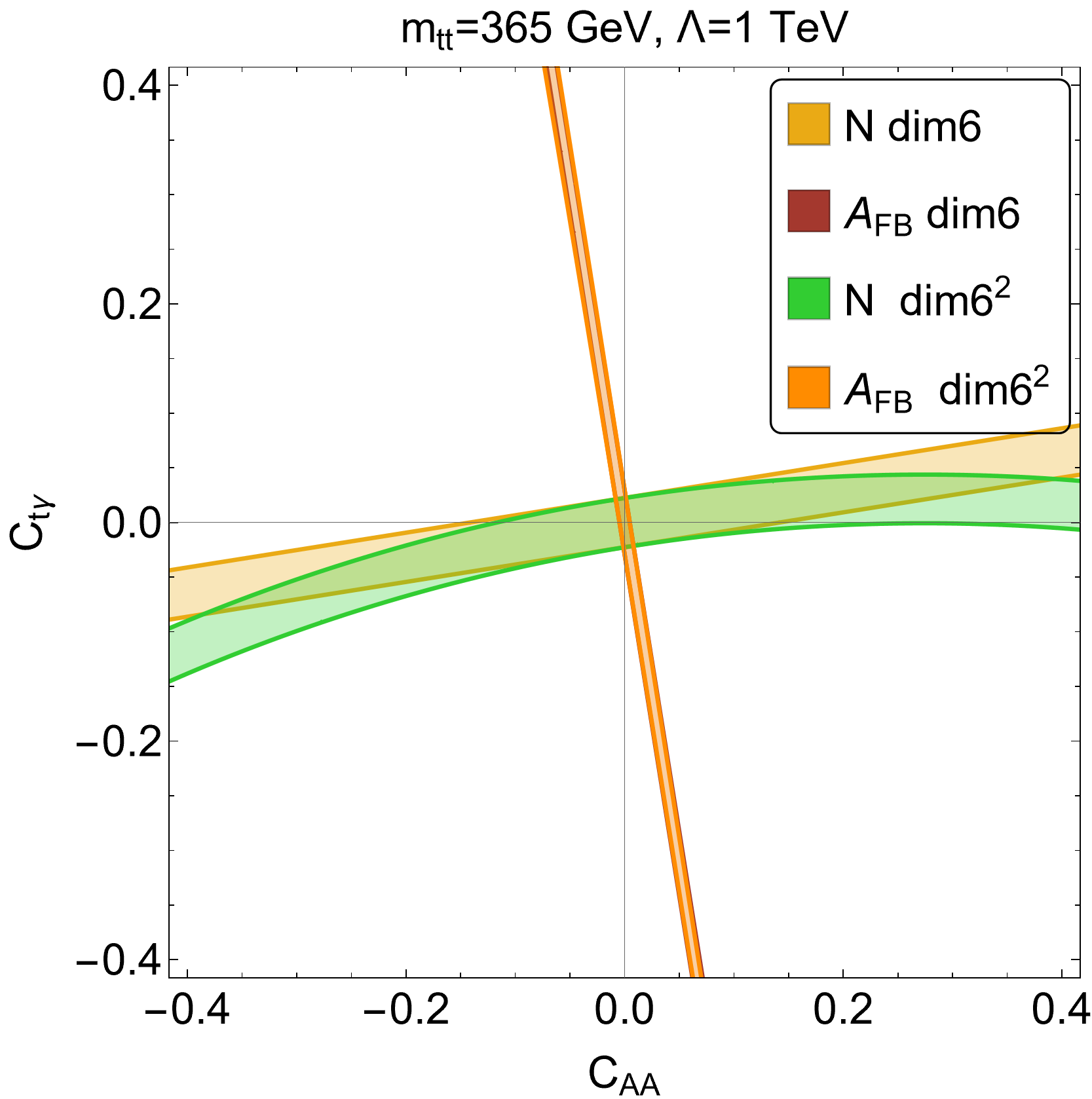}
 \end{subfigure}
     \caption{Same as in the Fig.~\ref{fig:2wilson2} in Section~\ref{subsec:2d} of the main text, for the pairs $(C_{VA}, C_{t\gamma})$, 
    $(C_{AV}, C_{t\gamma})$, and $(C_{AA}, C_{t\gamma})$.}
     \label{fig:2wilson4}
 \end{figure}

\begin{figure}[t]
     \centering
 \begin{subfigure}{0.43\textwidth}
     \centering
     \includegraphics[width=\textwidth]{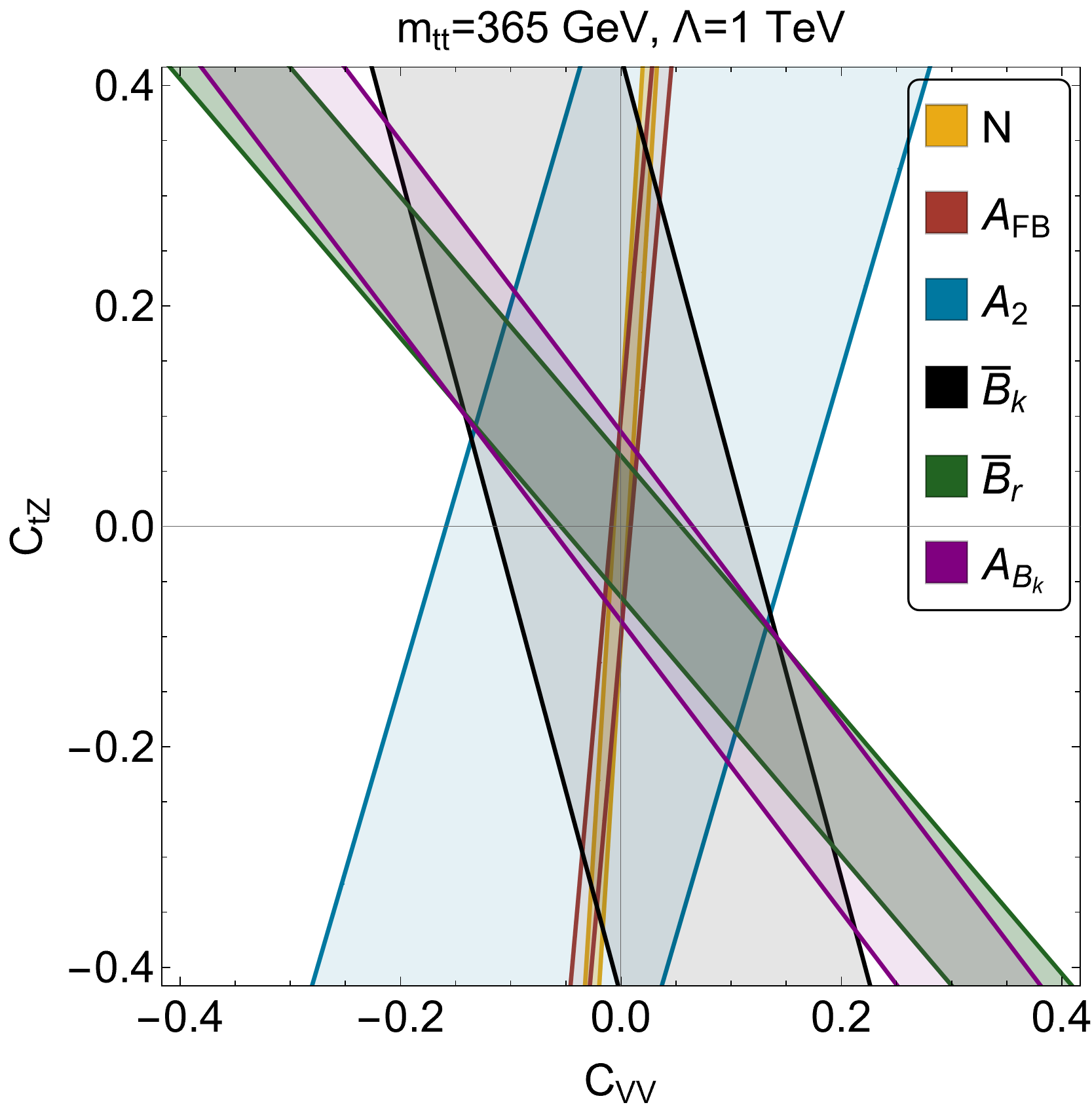}
 \end{subfigure}
  \begin{subfigure}{0.43\textwidth}
     \centering
     \includegraphics[width=\textwidth]{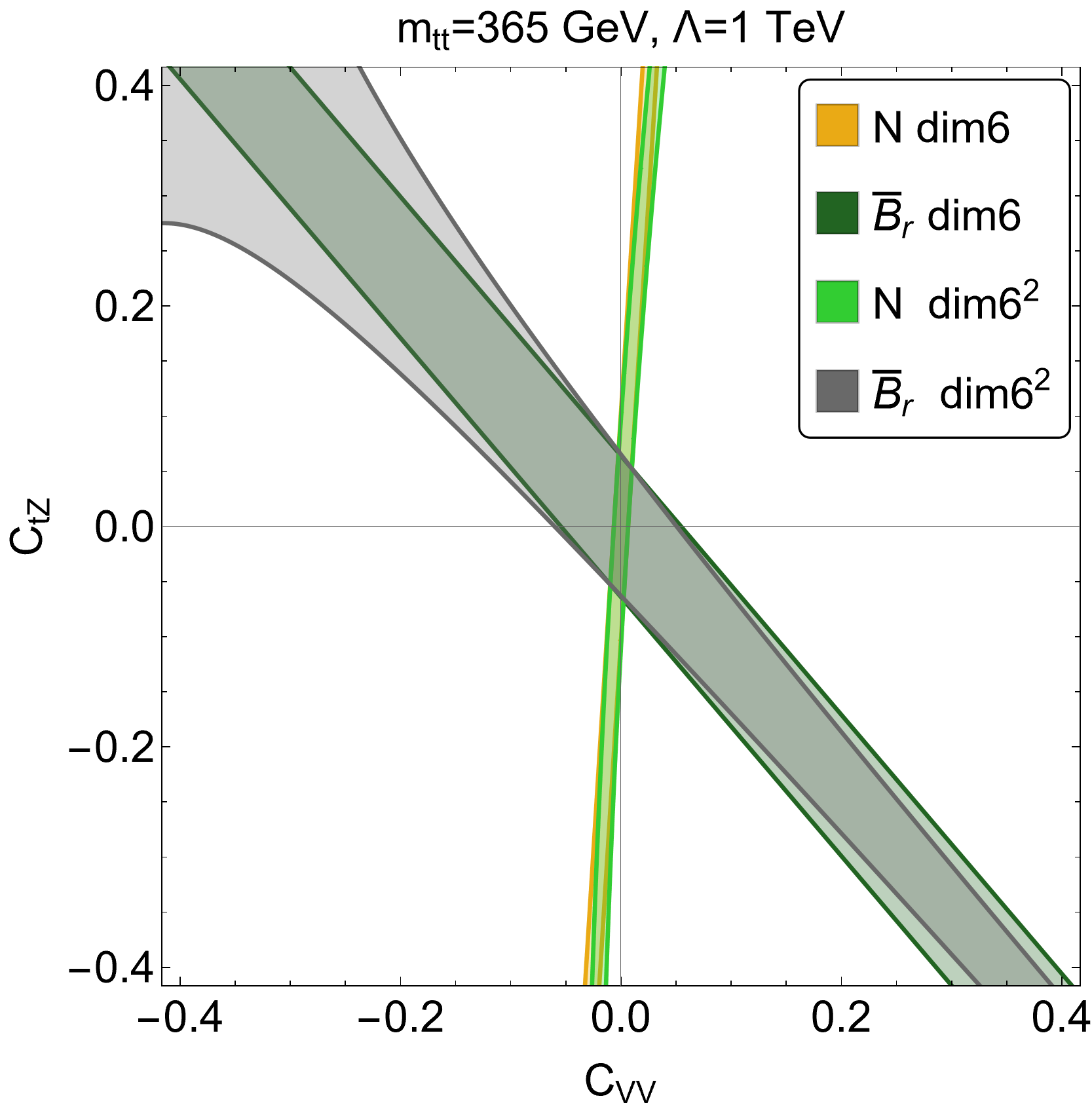}
 \end{subfigure}
  \begin{subfigure}{0.43\textwidth}
     \centering
     \includegraphics[width=\textwidth]{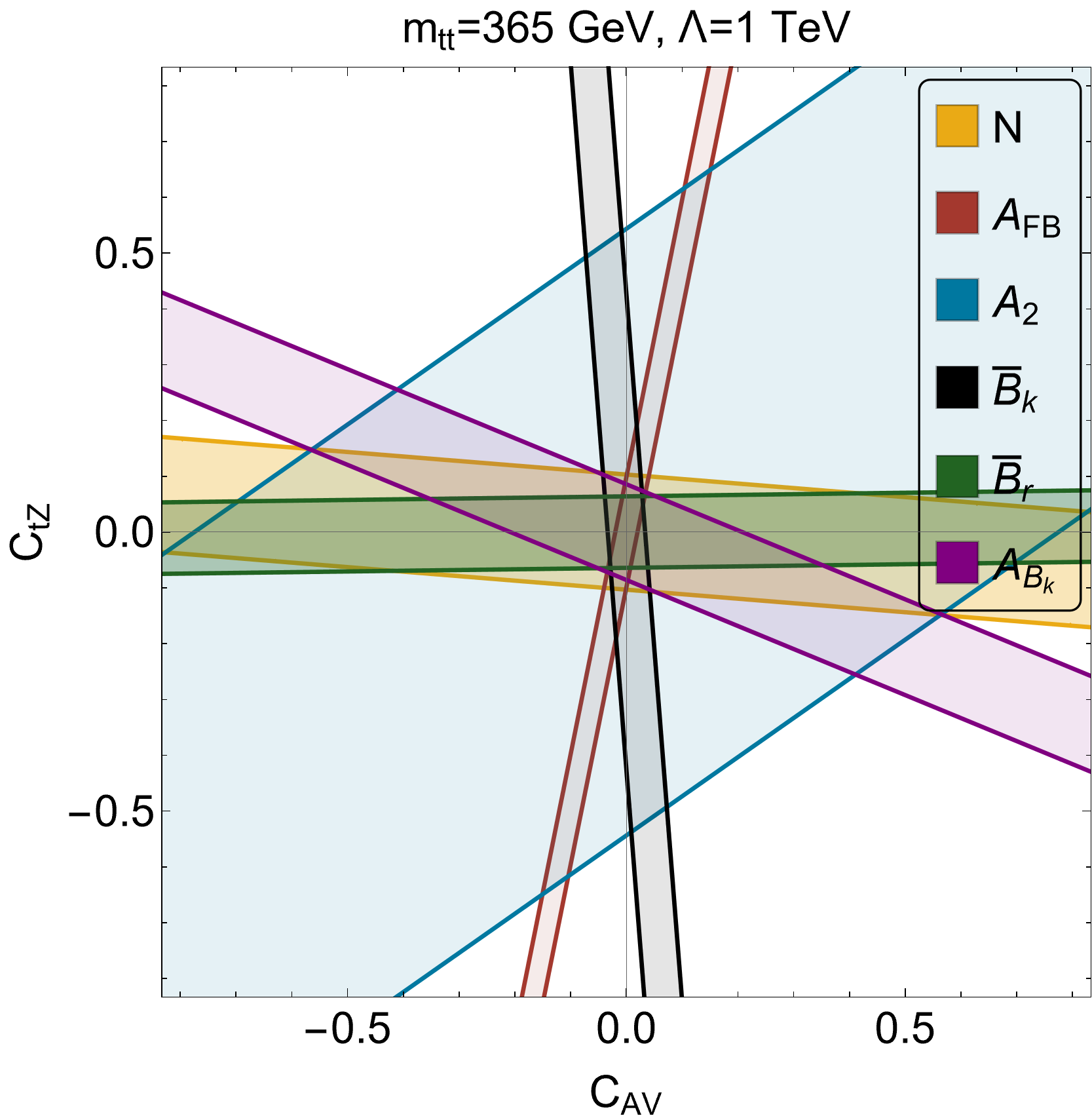}
 \end{subfigure}
  \begin{subfigure}{0.43\textwidth}
     \centering
     \includegraphics[width=\textwidth]{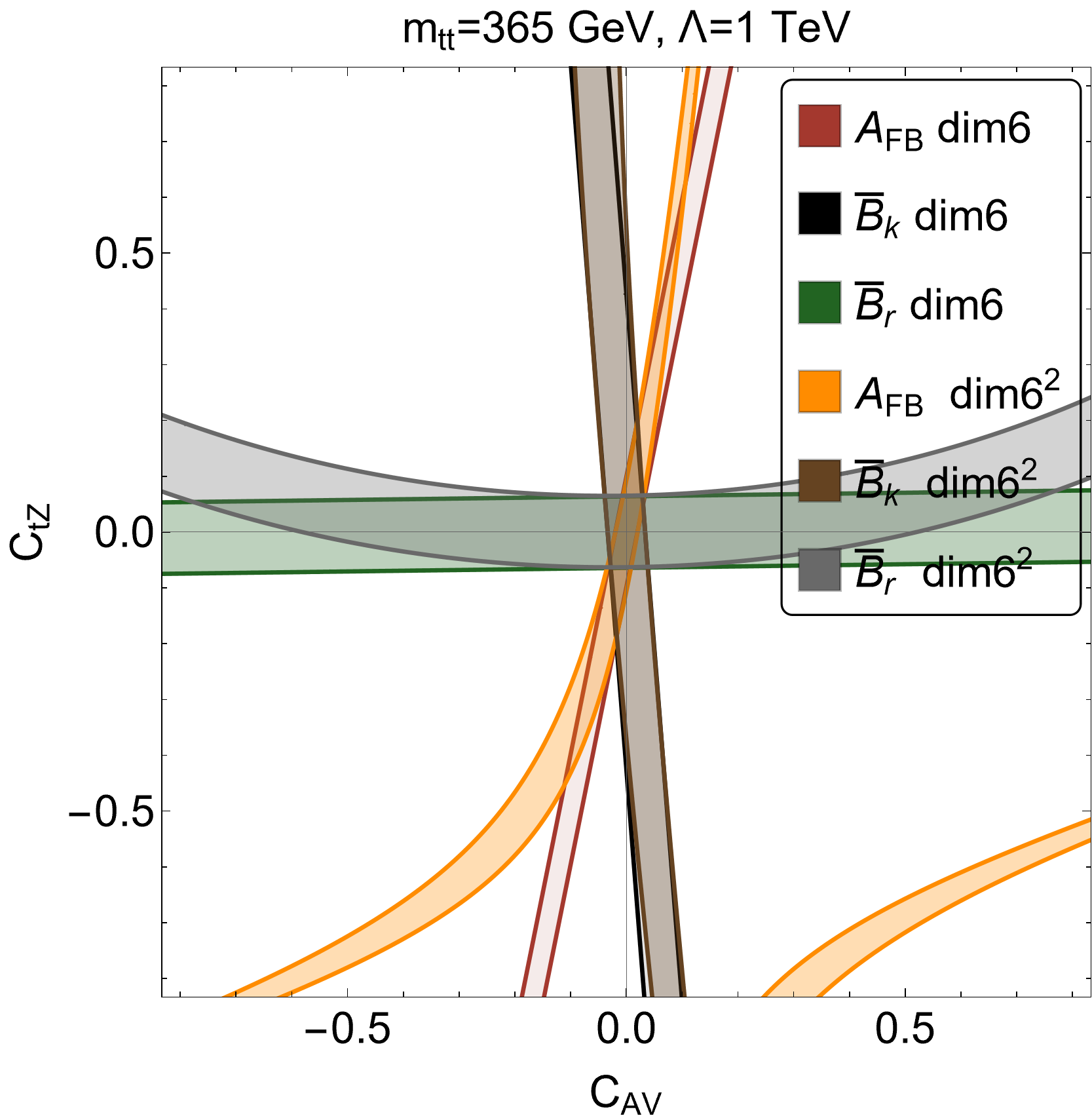}
 \end{subfigure}
    \begin{subfigure}{0.43\textwidth}
     \centering
     \includegraphics[width=\textwidth]{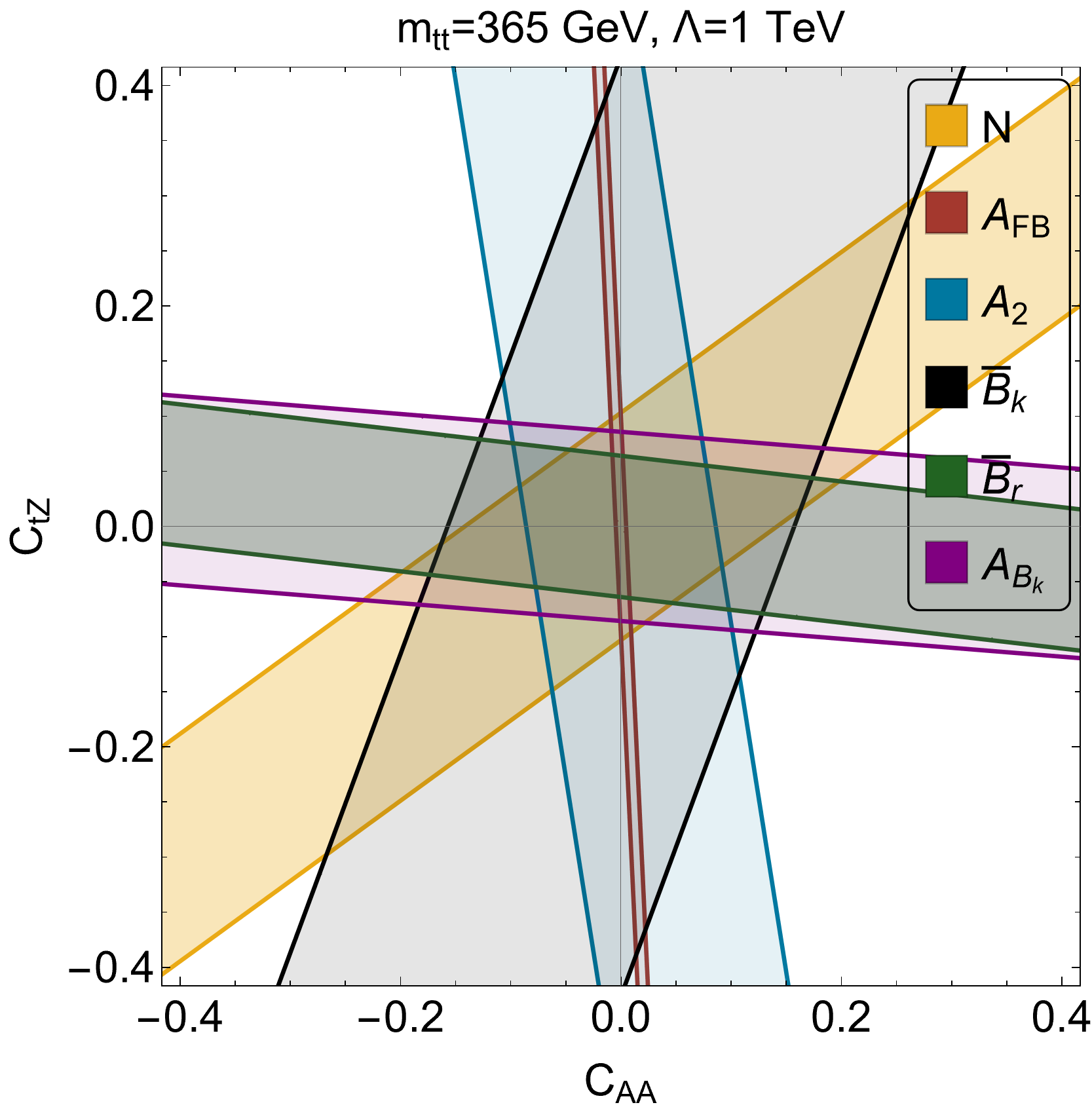}
 \end{subfigure}
 \begin{subfigure}{0.43\textwidth}
     \centering
     \includegraphics[width=\textwidth]{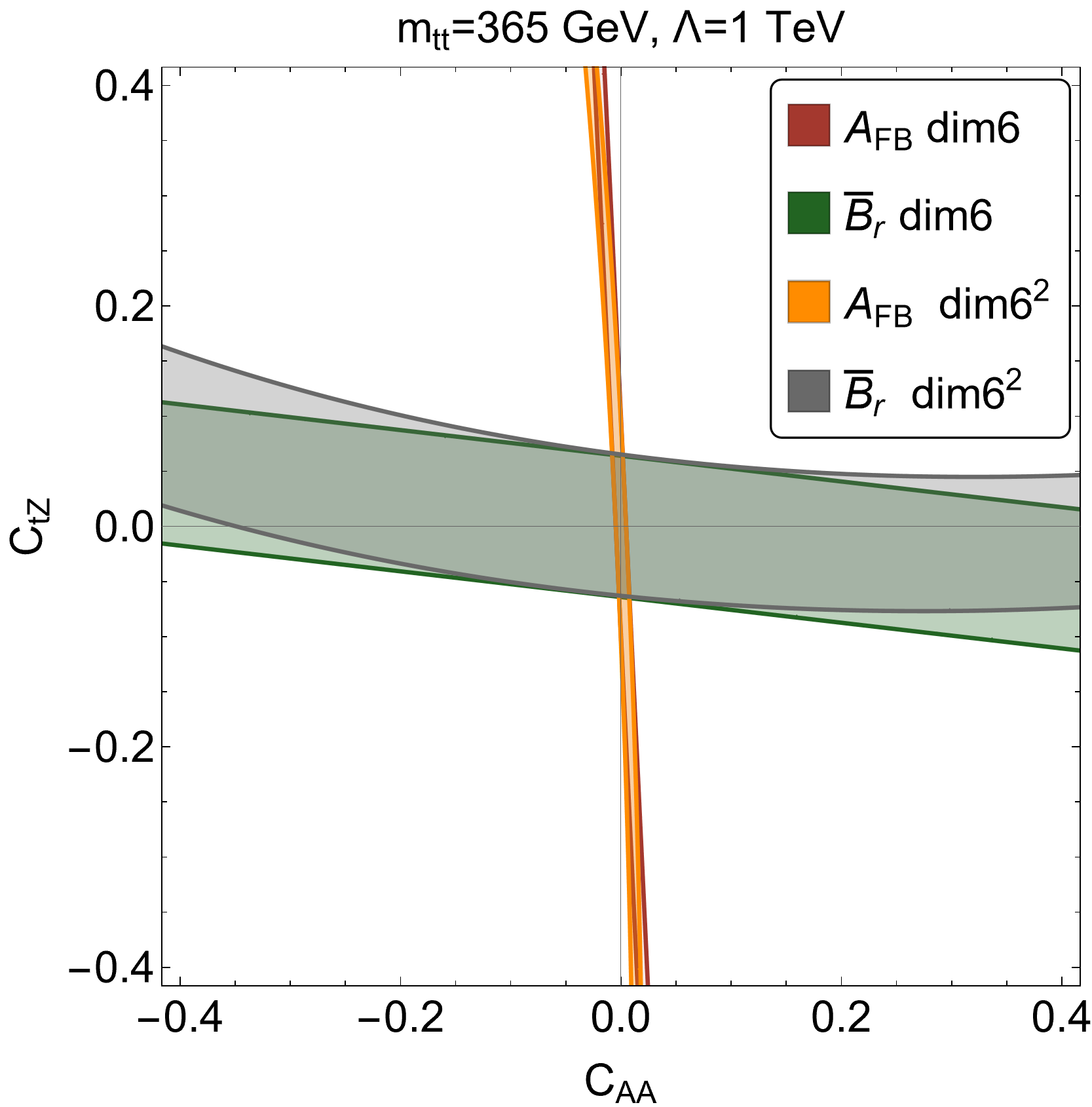}
 \end{subfigure}
     \caption{Same as in the Fig.~\ref{fig:2wilson2} in Section~\ref{subsec:2d} of the main text, for the pairs $(C_{VV}, C_{tZ})$, 
    $(C_{AV}, C_{tZ})$, and $(C_{AA}, C_{tZ})$.}
     \label{fig:2wilson5}
 \end{figure}

\begin{figure}[t]
     \centering
 \begin{subfigure}{0.43\textwidth}
     \centering
     \includegraphics[width=\textwidth]{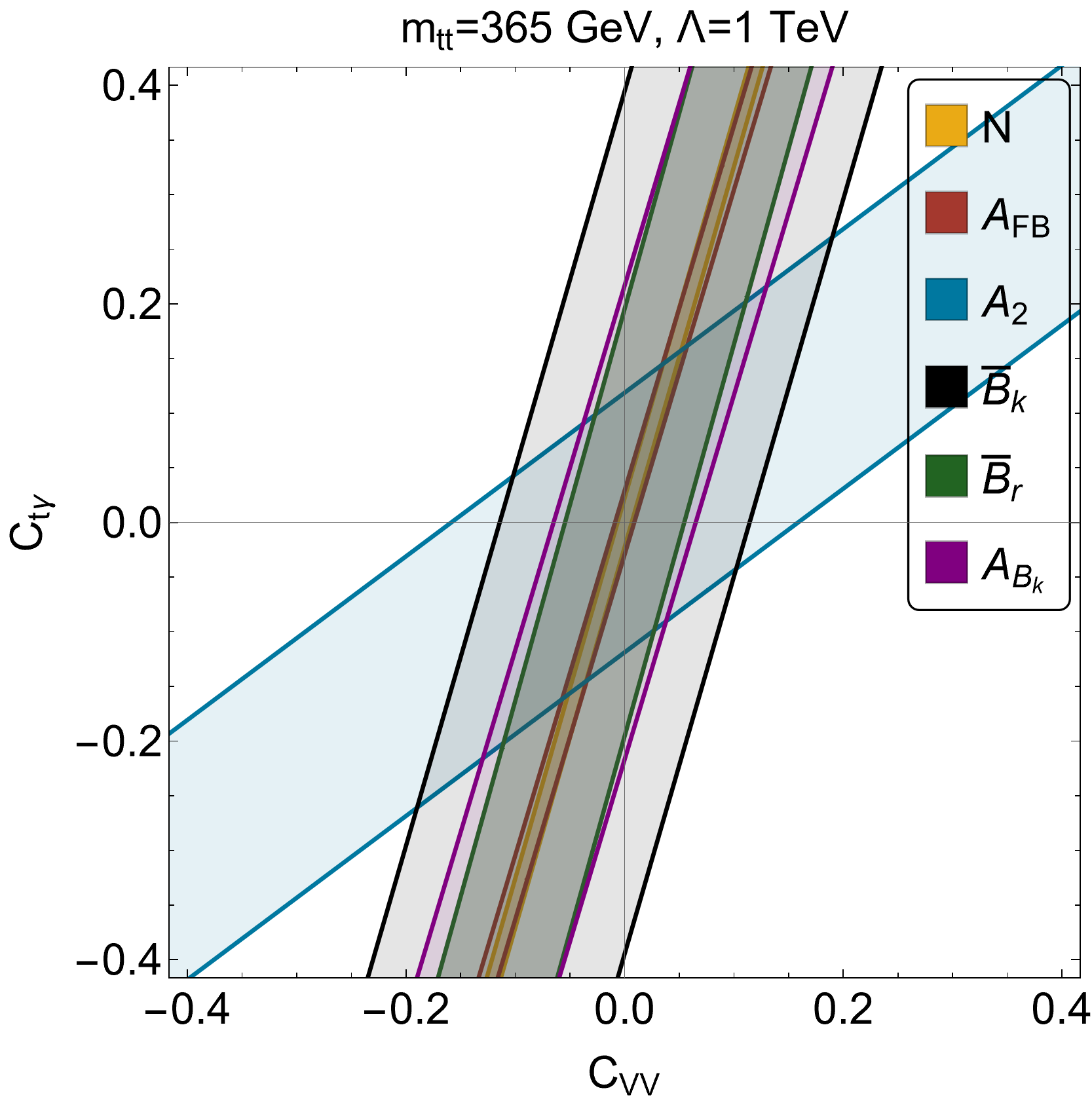}
 \end{subfigure}
  \begin{subfigure}{0.43\textwidth}
     \centering
     \includegraphics[width=\textwidth]{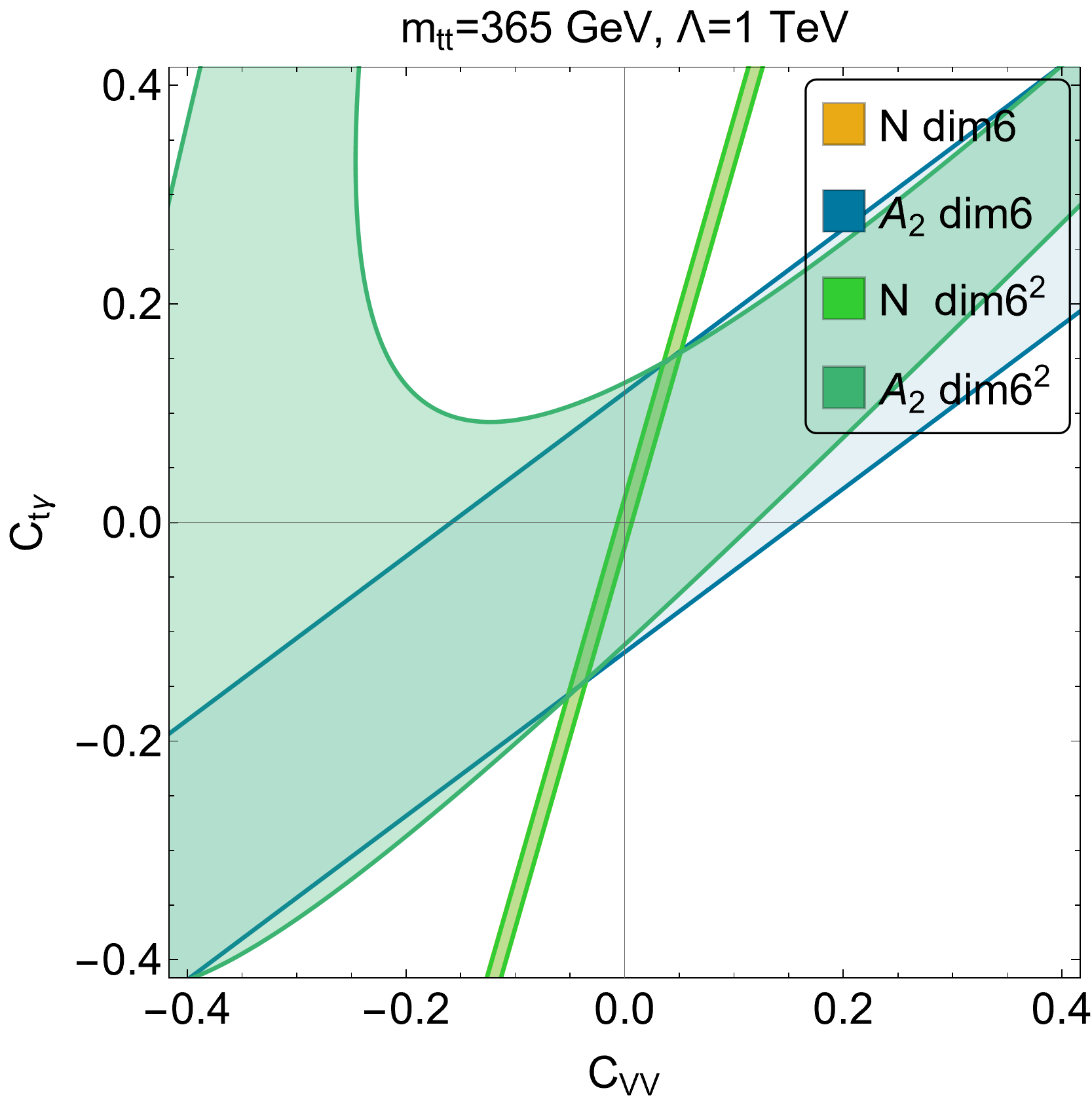}
 \end{subfigure}
  \begin{subfigure}{0.43\textwidth}
     \centering
     \includegraphics[width=\textwidth]{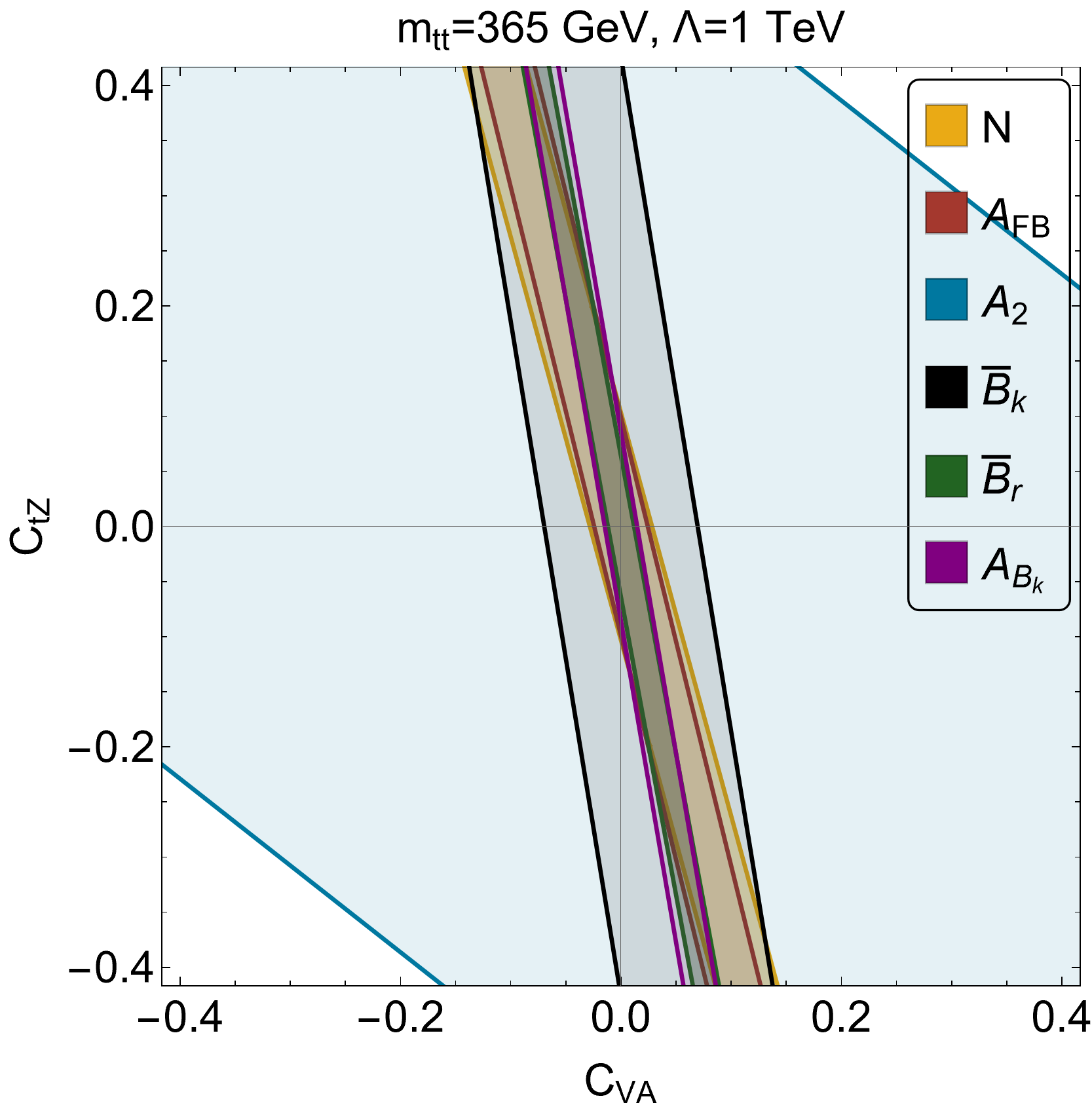}
 \end{subfigure}
  \begin{subfigure}{0.43\textwidth}
     \centering
     \includegraphics[width=\textwidth]{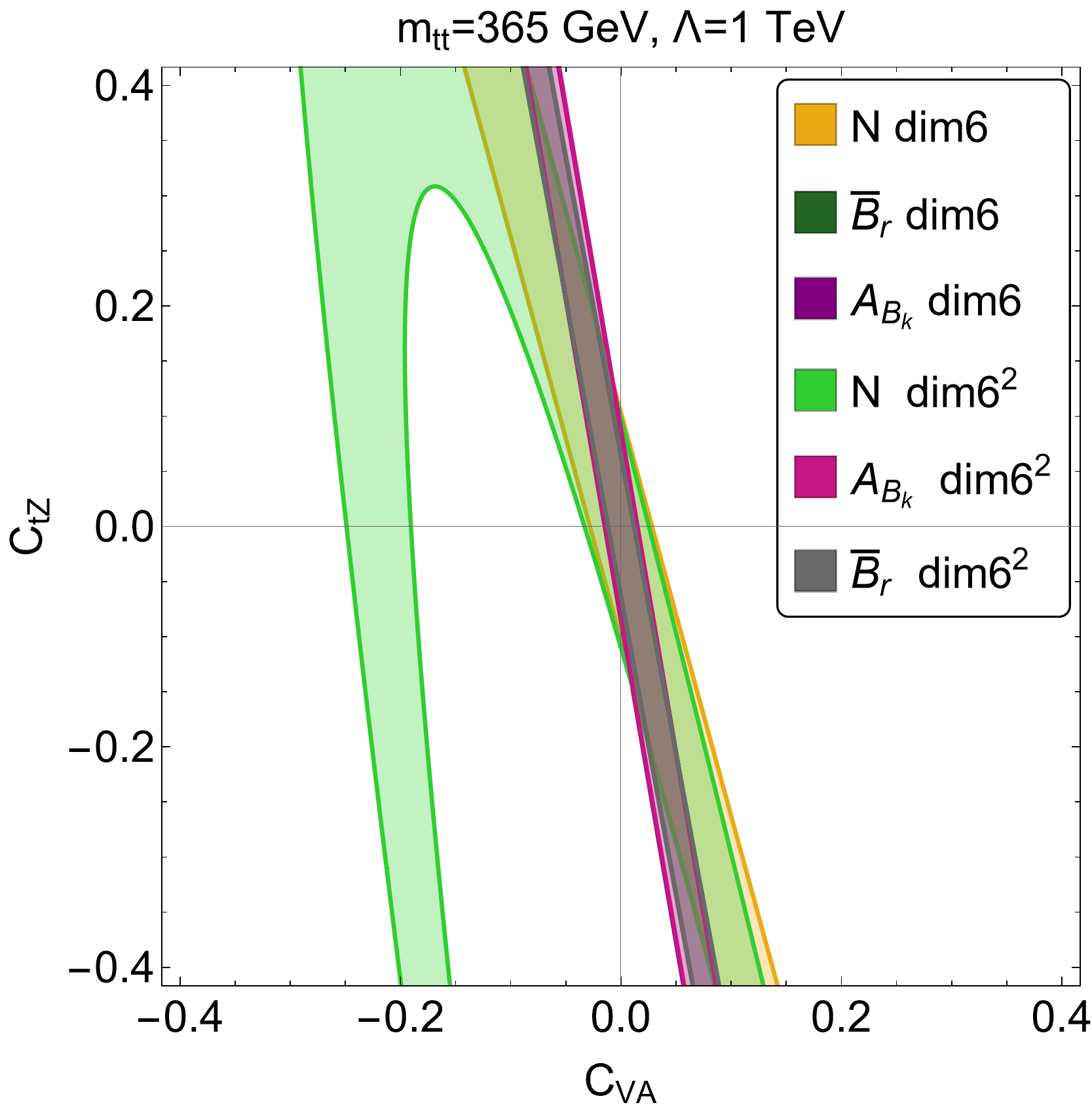}
 \end{subfigure}
     \caption{Same as in the Fig.~\ref{fig:2wilson2} in Section~\ref{subsec:2d} of the main text, for the pairs $(C_{VV}, C_{t\gamma})$, $(C_{VA}, C_{tZ})$.}
     \label{fig:2wilson6}
 \end{figure}

In this section, we collect all two-coefficient bounds for the simulated analysis
of Section~\ref{subsec:2d}, for $\sqrt{s}=365$ GeV in Figs.~\ref{fig:2wilson3},~\ref{fig:2wilson4}, ~\ref{fig:2wilson5}, and~\ref{fig:2wilson6}.

\end{document}